# Distinguishing Majorana zero modes from trivial defect states in an iron-based superconductor

Dongfei Wang[1,2,7] ✉, Jon Ortuzar[3,7], Freek Massee[4], Ruidan Zhong[5,6], Genda Gu[5], Wende Xiao[1], Yugui Yao[1], Roland Wiesendanger[2] ✉

[1] School of Physics; Centre for Quantum Physics, Key Laboratory of Advanced Optoelectronic Quantum Architecture and Measurement (MOE); State Key Laboratory of Chips and Systems for Advanced Light Field Display; Beijing Institute of Technology, 100081 Beijing, China

[2] Department of Physics, University of Hamburg, Jungiusstrasse 11, 20355 Hamburg, Germany

[3] CIC nanoGUNE-BRTA, 20018 Donostia-San Sebastián, Spain

[4] Université Paris-Saclay, CNRS, Laboratoire de Physique des Solides, 91405 Orsay, France.

[5] Condensed Matter Physics and Materials Science Department, Brookhaven National Laboratory, Upton, 11973 New York, USA

[6] Tsung-Dao Lee Institute, Shanghai Jiao Tong University, 201210 Shanghai, China

[7] These authors contributed equally: Dongfei Wang, Jon Ortuzar

✉ e-mails: dfwang@bit.edu.cn; wiesendanger@physnet.uni-hamburg.de

## Abstract

Majorana zero modes, which obey non-Abelian exchange statistics, are promising candidates for topological quantum computation due to their robustness against environmental perturbations. The iron-based superconductor Fe(Te,Se) has been identified as an intrinsic topological superconductor, possibly hosting Majorana zero modes. In this paper, we report the observation of near-zero-energy localized states at multiple structural defects on the Fe(Te,Se) surface, which could be misidentified as Majorana zero modes without additional verification. By using spin-polarized scanning tunneling spectroscopy, we demonstrate that the near-zero-energy localized states on step edges and line defects originate from topologically trivial Yu-Shiba-Rusinov states. In addition, zero-energy bound states are also observed for regions without surface defects. A combined spatial and magnetic field dependent analysis of the spin-resolved tunneling spectra in these regions reveals that this type of zero-energy states cannot be attributed to the presence of Majorana bound states. These findings emphasize the importance of spin-dependent studies of low-energy states for pursuing Majorana zero modes.

## Introduction

Quantum computing [1] emerges as a rapidly developing field because of its supremacy in dealing with problems like factoring large number [2] or database search [3] in a fast and efficient way. However, quantum bits (qubits) are usually very sensitive to the environment and easily lose their quantum coherence. Errors introduced by decoherence largely affect the reliability of the calculated results. Various algorithms have been proposed to correct for these errors [4-6], demonstrated by pioneering experiments [7,8]. As a drawback, significantly more qubits are needed to protect quantum information. On the other hand, topologically protected Majorana qubits [9] offer a route towards fault-tolerant quantum computation [10]. In this case, the quantum states are stored in a particle number space, which can only be changed when a topological transition occurs, being more robust against environmental perturbations. Moreover, Majorana anyons [11], which obey non-Abelian exchange statistics, were proposed as basic blocks to realize topological quantum computers [12], thereby hoping to improve the quantum fidelity in a more efficient way. However, until now, free Majorana anyons have not yet been detected.

In condensed matter, Majorana states emerge as quasi-particles, such as Majorana bound states (MBS), or dispersing Majorana modes (DMM), which were predicted to exist in various material platforms [13-20]. Numerous evidence for the existence of MBS or DMM has been reported in the last decade [21-29]. Although various hybrid material systems were demonstrated to host Majorana modes, there has been some reproducibility crisis [30] challenging this rapidly developing field. Intrinsic difficulties such as disorder in the device [31] or non-negligible contact with the substrate [32,33] make claims based on the measurement of a single physical quantity questionable. They do not disprove other sources with similar signatures. Pioneering experiments towards checking the zero energy bound states' nature by shot-noise [34] or spin-dependent measurements [35-37] have been reported, though studies in this direction are still rare due to the related experimental challenges.

Motivated by recent experiments indicating a possible topologically trivial origin of the low energy bound state [38,39] or dispersing modes [40,41] associated with defects of the topological superconductor Fe(Te,Se), we performed spin-polarized scanning tunneling microscopy/spectroscopy (SP-STM/STS) measurements of the bound states near various defects to study their topological nature. Our results suggest that the bound states near step edges and line defects most likely arise from topologically trivial Yu-Shiba-Rusinov (YSR) bound states [42-44] or YSR band formation [45,46] in superconducting Fe(Te,Se). These findings clearly demonstrate the importance of revealing the spin-dependent properties of zero-energy states as candidates for Majorana modes.

## Results

*Three types of defects at the surface of Fe(Te,Se)*

Freshly cleaved samples with stoichiometric ratio $FeTe_{0.55}Se_{0.45}$ in ultra-high vacuum usually exhibit three types of defects, namely magnetic impurities, line defects and step edges, as illustrated in Fig. 1a, e, i. The magnetic impurities are due to the excess Fe atoms resulting from the Fe(Te,Se) sample synthesis. They are distributed either on the cleaved surface or in the bulk as shown in the sketch in Fig. 1a,b. These magnetic Fe impurities can introduce near zero-energy bound states due to the exchange coupling with the Cooper pairs of the superconducting Fe(Te,Se). The energy position, spin polarization as well as the topological nature of such bound states have already been investigated for Fe adatoms on the Fe(Te,Se) surface [36,47,48]. In this study, we focus on bulk but near surface Fe impurities with bound states that can still be accessed by means of STS, as schematically shown in Fig. 1b. As an example, in Fig. 1d, we observe enhanced zero-energy density of states (DOS) at various positions, without the presence of surface defects (see Fig. 1c).

The second type of defect is given by line protrusions on the Fe(Te,Se) surface. There are two types of line defects observed: Type I is referred to the strain induced wrinkles, as schematically shown on the left of Fig. 1f. Their formations originate either from the sample preparation or sample cleavage process. In the STM topography image of Fig. 1g, two such wrinkles are revealed (see also Supplementary Fig. S1c-e). On the other hand, a Type II line defect with a bend can be also found as shown in Fig. 1h (details in Supplementary Fig. S1f-h). This type of defect has a narrower width but bigger height compared to the first type (see Supplementary Table S2), indicating a possible different formation mechanism. Interestingly, an enhanced near-zero DOS was observed for both formations as discussed later.

The third type of defects we studied are surface step edges. They are less commonly observed due to the high quality of the Fe(Te,Se) samples. In the STM topography image of Fig. 1k (and Supplementary Fig. S1a,b) we can see a step edge that is not atomically straight and exhibits a height difference of 1.23 nm, consisting of two triple-layers [49,50]. An enhanced near-zero DOS was also observed at the step edges, but with a very inhomogeneous spatial distribution, as can be seen in Fig. 1l. This finding can be explained latterly by the presence of bound states related to magnetic defect sites. Thus, we propose a step edge model as shown in Fig. 1j in which some edge atoms are missing leaving unpaired electrons at the edges, which can serve as magnetic scattering centers.

*Near-zero energy bound states at the edges*

We first focus on surface step edges as the one shown in Fig. 2a. We took spatially resolved STS data starting from the step edge along paths as indicated by the two 12 nm-long arrows. Close to the edge, path A shows two particle-hole symmetric peaks while path B reveals a zero-bias peak (ZBP) inside the superconducting gap, see Fig. 2b. From the STS line profiles shown in Fig. 2c, we can see that the two particle-hole symmetric peaks stay at fixed energy positions but fade away in intensity within 10 nm away from the step edge (more data shown in Supplementary Fig. S12). The spatial-dependent behavior could be explained by a YSR bound state originating from unpaired electrons at the step edge [51], consistent with our previous observation for surface Fe impurities [36]. On the other hand, the ZBP in line B as if splits within a few nanometers, as highlighted in the first 6 spectra in Fig. 2d. Similar to the peaks in path A, within 8 nm the splitted peaks vanish, see Fig. 2e. This behavior cannot be explained by a MBS, but rather by two interacting YSR bound states. It is well known that the interaction between impurities on a superconductor depends on the inter-impurity distance and has an impact on the measured spectra intensity distribution [52-54]. The shift in the position of the ZBP measured in Fig. 2d could be explained by the spatially dependent overlap of the wavefunction of two YSR bound states.

Qualitative agreement between the experimental data and a toy model simulation is obtained by assuming that the ZBP observed at the step edge results from a summation of two YSR peaks with similar peak heights plus a temperature broadening effect (see Fig. 2f , Supplementary Note 3 and Supplementary Fig. S2). To get the best fit, we used two different coupling strengths for spectra 1-2 and spectra 3-6. It implies more than one impurity contributed to the formation of the dispersing YSR bound state. Based on this finding, we constructed a more realistic model with two impurities which can reproduce the YSR intensity spatial distribution nicely (Supplementary Note 3 and Supplementary Fig. S3). Thus, the enhanced DOS at zero energy shown in Fig. 1l most probably originates from this type of YSR bound state related with unpaired electrons at step edge defects [51].

Other possibilities, such as topological superconducting edge states, can be ruled out since we observe bound state peaks instead of an enhanced DOS within the whole superconducting gap [55]. The observed edge states can also hardly be Andreev bound states because: 1) the Fe(Te,Se) is an $s\pm$-wave superconductor with no quasiparticles inside the gap causing Andreev reflection; 2) Andreev bound states require two reflection walls. Near the step edges shown in Fig. 2a, we did not observe a second wall leading to Andreev reflection. It is also unlikely to be composed of several close packed peaks with one residing at zero-energy as reported previously [56], as the observed splitting involves a spectral weight transfer from negative energy to positive energy which is not the case for a MBS.

*Enhanced in-gap density of state at two kinds of line defects*

Next, we examine the Type I line defect-related in-gap states located at the wrinkle (Fig. 1g). From the spectroscopic dI/dV maps, we can see a spatially inhomogeneous enhanced zero-energy DOS, as shown in Fig. 3a,b. For some specific position, as indicated by the white box in Fig. 3a, the zero-energy DOS is very pronounced and highly localized, as shown in Fig. 3b. In contrast, for other positions along the wrinkle there is no zero-energy DOS enhancement, as indicated by the white arrow. This observation clearly rules out the existence of a topological dispersing mode. To investigate the origin of this state, we focus on the region within the white box in more detail. From the small scale STM image of Fig. 3c we can see that there are no surface adatoms on top of or near the wrinkle. By analyzing the spatially dependent STS data across the wrinkle, we find clearly two peaks with symmetrical energy relative to zero inside the superconducting gap as seen in Fig. 3d. This observation is compatible with the presence of a YSR bound state.

As shown in Fig. 3e, we also observed a Type II line defect featured by a bend (Fig. 1h). Based on zero-energy dI/dV maps (Supplementary Fig. S4a, b), we noticed that the zero-energy DOS is enhanced at the bend, while there is almost no enhancement for nearby positions. To get a deeper understanding of this type of defect state, we took line STS data across and along the bend, as indicated in Fig. 3f,h. The results are shown in Fig. 3g,i correspondingly. First, we notice that the superconducting gap gets filled at the bend but recovers to the bulk value 4 nm away from it. Second, it is hard to conclude any systematic spatial dependence in the STS data shown in Fig. 3i for the bend, indicating that the defect states are quite inhomogeneous. Similar results have been reported before [57], interpreting the flat line-shape DOS enhancement as dispersing Majorana mode caused by a $\pi$ phase shift of the lattice (right of Fig. 1f) near a 1D defect mimicking the Fu-Kane model [13]. Recently, however, using a more refined method to extract the phase shift, it was argued that the phase shift of this 1D defect is not $\pi$, but $\pi/2$, and additional measurements further highlighted their topological trivial character [41]. Thus, it is important to accurately investigate the phase shift as well as the topological property of the bend.

*Spin-polarized dispersing mode at Type II line defect*

Based on atomic-resolution STM data near the bend shown in Fig. 4a (see also Supplementary Fig. S5a, d and e), it is possible to extract the lattice shift between the two regions on the left and right. Figure 4b (Fast Fourier Transform (FFT) map of Fig. 4a) reveals four peaks related to the square lattice of the Fe(Te,Se) surface with Bragg peak 1 showing a finite splitting. The splitting of Bragg peak 2 is hard to see. To get the exact phase shift of this region, we employed the same method as discussed in Ref. [41]. As seen in Fig. 4c,d, there is indeed a phase difference for the first Bragg peak, while much less for the second. We took a linecut, as indicated by the dashed line, to extract the exact phase values across the bend (Supplementary Fig. S6a). By plotting all the phase values, we got a statistical diagram (Supplementary Fig. S6b) indicating that there is almost no phase shift for the second Bragg peak and a phase shift near 0.5 $\pi$ for the first.

As the phase difference we observe here is much less than $\pi$, we seek for other explanations for the enhanced DOS at the line defect bend beyond a Majorana mode by measuring the magnetic field response of the tunneling spectra within $\pm$ 2 mV. We selected three positions labeled 1-3 with atomic precision (Fig. 4e) and obtained the corresponding SP-STS results (Fig. 4f). First, we polarized the STM tip in a +1 T field and measured the SP-STS spectra at the three different positions. Then, we switched the tip's spin direction in a -1 T field and measured the corresponding SP-STS spectra again. To check the reproducibility of our data, we polarized the tip again at +1 T and measured the SP-STS spectra. (The detailed tip calibration before the SP-STS measurement is presented in Supplementary Fig. S7). For the tip being polarized at +1 T and -1 T, we can clearly see a DOS difference within the energy range $\pm$ 0.8 mV, beyond which quasiparticles near the coherence peaks come into play. More importantly, we find that the sign of the polarization of the DOS is reversed for positive and negative energies. This observation is similar to the spin-polarization of YSR bound states, which is also of opposite sign for positive and negative energies [36]. Therefore, we propose that the enhanced DOS within the superconducting gap may be caused by the interactions between several magnetic impurities. The most plausible origin for the magnetic impurities is the excess Fe atoms introduced during the sample synthesis. The observation of surface Fe adatoms has been reported repeatedly for this type of sample with the same stoichiometric ratio.

There is additional evidence for this hypothesis: A careful identification of the crossing point between the SP-STS spectra at +1 T and -1 T reveals that it is not at zero energy for position 1 and position 2. By modeling two ferromagnetically coupled magnetic impurities with proper YSR bound state splitting energy (see Fig. 4g), we can see that the spin-polarization is already not particle-hole symmetric. Therefore, the defect states observed at the bend are most probably induced by some coupled magnetic impurities below the surface which contribute to the DOS inside the superconducting gap. Figure 4h shows the model structure simulating the defect with an ensemble of magnetic impurities coupled in line but irregularly (Supplementary Note 3). The large ensemble of impurities generates a DOS that fills the superconducting gap of Fe(Te,Se) (see Supplementary Fig. S10), which is in agreement with the spectra shown in Fig. 3g,i. Moreover, the complex overlaps between the wavefunction of the individual YSR states generated asymmetric SP-STS. The example in two points of the ensemble is given in Fig. 4i, which qualitatively simulates the measured SP-STS in Fig. 4f.

*Trivial bound states below the surface*

At last, we discuss the surprisingly enhanced zero-energy DOS in regions with no surface defects. As shown in Fig. 5a, with atomic resolution, we did not observe any adatom or line defect within this region. However, the zero-energy dI/dV map of the same region shows a donut-shaped signal enhancement with two local maxima labeled as point 1 and point 2 in Fig. 5b. At zero magnetic field, the STS data for these two points (Fig. 5c) show both zero bias peaks with the same Full-Width at Half-Maximum (HWHM) near 0.5 mV, which is comparable to the ZBP observed in vortex cores at lower temperature [58]. To get a deeper understanding of this zero-bias peak, we took spatially as well as magnetic-field dependent STS data. STS spectra are taken in three directions, indicated by arrows A to C in Fig. 5d-i. There is no clear energy shift of the zero-bias peak within lines A and B. But a finite shift to positive energies of the peak in line C can be observed (see also Supplementary Fig. S9). The peak energy position along the lines can be better seen in Fig. 5j. As MBS should not shift in peak energy, the bound state at point 1 is probably caused by some other mechanism. For some magnetic impurities below the surface, a spatially dispersing YSR bound state was reported caused by a tip gating effect [38]. On the other hand, the non-dispersing ZBP across point 2 could be an indication for an MBS. As a MBS is spin-polarized [59], we measured SP-STS spectra for two positions with a calibrated Cr tip (see Supplementary Fig. S8). We can clearly see in Fig. 5k,l that the ZBP intensity at point 1 exhibits a magnetic response, while this is not obvious for point 2 (see also Supplementary Note 3 and Supplementary Fig. S11 for the model fit). Thus, the ZBP at point 2 may not be caused by a spin-polarized MBS. The ZBP near point 2 does not show spatial dispersion, most probably because the defect is located further away from the surface compared to point 1, where the tip gating effect is quite small. This is supported by a lower zero-energy DOS for point 2 as shown in Fig. 5c. However, we also realized that when the defect is more far away from the surface, the spin sensitivity of the measurement can be reduced. Thus, other possible explanations for the zero-energy bound state at point 2 may exist.

## Discussion

In conclusion, we observed a zero-energy DOS enhancement at three different types of defects near the surface of Fe(Te,Se), namely impurities below the surface, step edges and line defects. By using spatially dependent as well as spin-polarized STS, a ZBP splitting was observed at the step edges which can be modeled by YSR states mixed with different peak intensities at finite temperature. The enhanced zero-energy DOS on the wrinkle is mostly due to several distributed magnetic impurities. On the line defect bend, the flat DOS within the superconducting gap is most likely due to irregularly coupled magnetic impurities instead of a dispersing Majorana mode with topological origin. Lastly, we observed ZBPs in regions with no surface defects. By combining spatially dependent and magnetic field dependent SP-STS measurements, we found that none of them can be explained by MBS. Our results emphasize the importance of using multiple methods to examine the nature of ZBP.

## Methods

The Fe(Te,Se) samples we used in the experiment were grown using the self-flux method with a final Fe, Te, Se concentration of $FeTe_{0.55}Se_{0.45}$. The superconducting transition temperature $T_c$ of the sample is 14.5 K. With no further degassing, the Fe(Te,Se) crystals were cleaved in the preparation chamber by vacuum tape with a vacuum better than $5 \times 10^{-10}$ mbar. Then the samples were transferred *in-situ* to the low-temperature ultra-high vacuum (UHV) STM. All STM and STS measurements were performed in the UHV-STM chamber with a vacuum better than $2 \times 10^{-10}$ at a temperature of 1.1 K. Magnetic fields up to 3 T perpendicular to the sample surface could be applied. Differential tunneling conductance (dI/dV) spectra were recorded using a lock-in technique with a modulation bias $V_{bias}$ and a modulation frequency near 900 Hz. The STM tip was stabilized first at a current ($I_{stab}$) for a while, then the feedback was switched off to record the spectra. For most of our measurements, we used a tunneling conductance on the order of $10^{-4}\ 2e^2/h$. We employed a bulk Cr tip as our sensor to perform SP-STM measurements. To switch the spin direction of the antiferromagnetic Cr tip in an external field up to 3 T, we dipped our tip into an Fe film on a W(110) substrate, leaving a ferromagnetic Fe coating at the tip apex. Details about the tip's spin-polarization calibration can be found in the Supplementary Information.

## Data Availability

Relevant data supporting the key findings of this study are available within the article and in the Supplementary Information/Source data file. Additional data generated during the current study are available from the corresponding authors upon request. Source data are provided with this paper.

## References

1 Nielsen, M. A. & Chuang, I. L. *Quantum computation and quantum information*. (Cambridge university press, 2010).

2 Shor, P. W. Algorithms for quantum computation: discrete logarithms and factoring. *Proceedings 35th annual symposium on foundations of computer science*, 124-134 (1994).

3 Grover, L. K. A fast quantum mechanical algorithm for database search. *Proceedings of the twenty-eighth annual ACM symposium on theory of computing*, 212-219 (1996).

4 Calderbank, A. R. & Shor, P. W. Good quantum error-correcting codes exist. *Phys. Rev. A* **54**, 1098 (1996).

5 Kitaev, A. Y. Quantum computations: algorithms and error correction. *Russ. Math. Surv.* **52**, 1191 (1997).

6 Braunstein, S. L. Quantum error correction for communication with linear optics. *Nature* **394**, 47-49 (1998).

7 Chiaverini, J. *et al.* Realization of quantum error correction. *Nature* **432**, 602-605 (2004).

8 Kelly, J. *et al.* State preservation by repetitive error detection in a superconducting quantum circuit. *Nature* **519**, 66-69 (2015).

9 Nayak, C., Simon, S. H., Stern, A., Freedman, M. & Das Sarma, S. Non-Abelian anyons and topological quantum computation. *Rev. Mod. Phys.* **80**, 1083-1159 (2008).

10 Kitaev, A. Y. Fault-tolerant quantum computation by anyons. *Ann. Phys.* **303**, 2-30 (2003).

11 Kitaev, A. Y. Unpaired Majorana fermions in quantum wires. *Phys.-Uspekhi* **44**, 131 (2001).

12 Sarma, S. D., Freedman, M. & Nayak, C. Majorana zero modes and topological quantum computation. *npj Quantum Inf.* **1**, 1-13 (2015).

13 Fu, L. & Kane, C. Superconducting proximity effect and majorana fermions at the surface of a topological insulator. *Phys. Rev. Lett.* **100**, 096407 (2008).

14 Sau, J. D., Lutchyn, R. M., Tewari, S. & Sarma, S. D. Generic new platform for topological quantum computation using semiconductor heterostructures. *Phys. Rev. Lett.* **104**, 040502 (2010).

15 Lutchyn, R. M., Sau, J. D. & Das Sarma, S. Majorana Fermions and a Topological Phase Transition in Semiconductor-Superconductor Heterostructures. *Phys. Rev. Lett.* **105**, 077001 (2010).

16 Braunecker, B. & Simon, P. Interplay between classical magnetic moments and superconductivity in quantum one-dimensional conductors: toward a self-sustained topological Majorana phase. *Phys. Rev. Lett.* **111**, 147202 (2013).

17 Yang, G., Stano, P., Klinovaja, J. & Loss, D. Majorana bound states in magnetic skyrmions. *Phys. Rev. B* **93**, 224505 (2016).

18 Pershoguba, S. S., Nakosai, S. & Balatsky, A. V. Skyrmion-induced bound states in a superconductor. *Phys. Rev. B* **94**, 064513 (2016).

19 Li, J. *et al.* Two-dimensional chiral topological superconductivity in Shiba lattices. *Nat. Commun.* **7**, 12297 (2016).

20 Pientka, F. *et al.* Topological superconductivity in a planar Josephson junction. *Phys. Rev. X* **7**, 021032 (2017).

21 Mourik, V. *et al.* Signatures of Majorana fermions in hybrid superconductor-semiconductor nanowire devices. *Science* **336**, 1003-1007 (2012).

22 Deng, M. *et al.* Majorana bound state in a coupled quantum-dot hybrid-nanowire system. *Science* **354**, 1557-1562 (2016).

23 Dvir, T. *et al.* Realization of a minimal Kitaev chain in coupled quantum dots. *Nature* **614**, 445-450 (2023).

24 Nadj-Perge, S. *et al.* Observation of Majorana fermions in ferromagnetic atomic chains on a superconductor. *Science* **346**, 602-607 (2014).

25 Xu, J.-P. *et al.* Experimental detection of a Majorana mode in the core of a magnetic vortex inside a topological insulator-superconductor Bi 2 Te 3/NbSe 2 heterostructure. *Phys. Rev. Lett.* **114**, 017001 (2015).

26 Fornieri, A. *et al.* Evidence of topological superconductivity in planar Josephson junctions. *Nature* **569**, 89-92 (2019).

27 Ren, H. *et al.* Topological superconductivity in a phase-controlled Josephson junction. *Nature*

**569**, 93-98 (2019).

28 Palacio-Morales, A. *et al.* Atomic-scale interface engineering of Majorana edge modes in a 2D magnet-superconductor hybrid system. *Sci. Adv.* **5**, eaav6600 (2019).

29 Schneider, L. *et al.* Precursors of Majorana modes and their length-dependent energy oscillations probed at both ends of atomic Shiba chains. *Nat. Nanotechnol.* **17**, 384-389 (2022).

30 Frolov, S. Quantum computing's reproducibility crisis: Majorana fermions. *Nature* **592**, 350-352 (2021).

31 Zhang, H. *et al.* Retraction note: Quantized majorana conductance. *Nature* **591**, E30 (2021).

32 Thorp, H. H. Editorial Retraction. *Science* **378**, 718-718 (2022).

33 Kayyalha, M. *et al.* Absence of evidence for chiral Majorana modes in quantum anomalous Hall-superconductor devices. *Science* **367**, 64-67 (2020).

34 Ge, J.-F. *et al.* Single-electron charge transfer into putative Majorana and trivial modes in individual vortices. *Nat. Commun.* **14**, 3341 (2023).

35 Jeon, S. *et al.* Distinguishing a Majorana zero mode using spin-resolved measurements. *Science* **358**, 772-776 (2017).

36 Wang, D., Wiebe, J., Zhong, R., Gu, G. & Wiesendanger, R. Spin-polarized Yu-Shiba-Rusinov states in an iron-based superconductor. *Phys. Rev. Lett.* **126**, 076802 (2021).

37 Ruby, M., Heinrich, B. W., Peng, Y., von Oppen, F. & Franke, K. J. Exploring a proximity-coupled Co chain on Pb (110) as a possible Majorana platform. *Nano lett.* **17**, 4473-4477 (2017).

38 Chatzopoulos, D. *et al.* Spatially dispersing Yu-Shiba-Rusinov states in the unconventional superconductor FeTe0. 55Se0. 45. *Nat. Commun.* **12**, 298 (2021).

39 Uldemolins, M. *et al.* Hund's coupling mediated multi-channel quantum phase transition of a single magnetic impurity in Fe (Se, Te). *Nat. Commun.* **15**, 8526 (2024).

40 Zhao, H. *et al.* Nematic transition and nanoscale suppression of superconductivity in Fe (Te, Se). *Nat. Phys.* **17**, 903-908 (2021).

41 Mesaros, A., Gu, G. & Massee, F. Topologically trivial gap-filling in superconducting Fe (Se, Te) by one-dimensional defects. *Nat. Commun.* **15**, 3774 (2024).

42 Luh, Y. Bound state in superconductors with paramagnetic impurities. *Acta Phys. Sin.* **21**, 75 (1965).

43 Shiba, H. Classical spins in superconductors. *Prog. Theor. Phys.* **40**, 435-451 (1968).

44 Rusinov, A. Theory of gapless superconductivity in alloys containing paramagnetic impurities. *Sov. Phys. JETP* **29**, 1101-1106 (1969).

45 Zhang, G. *et al.* Yu-Shiba-Rusinov bands in ferromagnetic superconducting diamond. *Sci. Adv.* **6**, eaaz2536 (2020).

46 Nadj-Perge, S., Drozdov, I. K., Bernevig, B. A. & Yazdani, A. Proposal for realizing Majorana fermions in chains of magnetic atoms on a superconductor. *Phys. Rev. B* **88**, 020407 (2013).

47 Yin, J.-X. *et al.* Observation of a robust zero-energy bound state in iron-based superconductor Fe (Te, Se). *Nat. Phys.* **11**, 543-546 (2015).

48 Fan, P. *et al.* Observation of magnetic adatom-induced Majorana vortex and its hybridization with field-induced Majorana vortex in an iron-based superconductor. *Nat. Commun.* **12**, 1348 (2021).

49 Chen, G. *et al.* Evidence for a robust sign-changing s-wave order parameter in monolayer films of superconducting Fe (Se, Te)/Bi2Te3. *npj Quantum Mater.* **7**, 110 (2022).

50 Gawryluk, D. *et al.* Growth conditions, structure and superconductivity of pure and metal-doped

FeTe1− xSex single crystals. *Supercond. Sci. Technol.* **24**, 065011 (2011).
51 Li, Y. *et al.* Observation of Yu-Shiba-Rusinov-like states at the edge of CrBr3/NbSe2 heterostructure. *Nat. Commun.* **15**, 10121 (2024).
52 Ruby, M., Heinrich, B. W., Peng, Y., von Oppen, F. & Franke, K. J. Wave-function hybridization in Yu-Shiba-Rusinov dimers. *Phys. Rev. Lett.* **120**, 156803 (2018).
53 Küster, F., Brinker, S., Lounis, S., Parkin, S. S. & Sessi, P. Long range and highly tunable interaction between local spins coupled to a superconducting condensate. *Nat. Commun.* **12**, 6722 (2021).
54 Ortuzar, J. *et al.* Yu-Shiba-Rusinov states in two-dimensional superconductors with arbitrary Fermi contours. *Phys. Rev. B* **105**, 245403 (2022).
55 Wang, Z. *et al.* Topological edge states in a high-temperature superconductor FeSe/SrTiO3 (001) film. *Nat. Mater.* **15**, 968-973 (2016).
56 Machida, T. *et al.* Zero-energy vortex bound state in the superconducting topological surface state of Fe (Se, Te). *Nat. Mater.* **18**, 811-815 (2019).
57 Wang, Z. *et al.* Evidence for dispersing 1D Majorana channels in an iron-based superconductor. *Science* **367**, 104-108 (2020).
58 Wang, D. *et al.* Evidence for Majorana bound states in an iron-based superconductor. *Science* **362**, 333-335 (2018).
59 Kawakami, T. & Hu, X. Evolution of density of states and a spin-resolved checkerboard-type pattern associated with the Majorana bound state. *Phys. Rev. Lett.* **115**, 177001 (2015).

## Acknowledgements

We would like to thank Jens Wiebe, De-Liang Bao, and Chuanchang Zeng for useful discussions as well as Torben Hänke and Anand Kamlapure for technical support. We also thank Hong Ding, Fazhi Yang and Cuihua Liu for providing the samples.

## Funding Statement

D.W., Y.Y. and W.X. gratefully acknowledge funding by National Natural Science Foundation of China (No. 12474474; No. 12321004; No. 12274029). R.W. and D.W. gratefully acknowledge funding by the EU via the ERC Advanced Grant ADMIRE (No. 786020), the DFG via the Cluster of Excellence "Advanced Imaging of Matter" (EXC 2056, project ID 390715994). F.M. acknowledges funding from the ANR (ANR-21-CE30-0017-01). R.Z. and G.G. acknowledges funding from the US Department of Energy, office of Basic Energy Sciences (contract no. de- sc0012704).

## Author Contributions

R.W. supervised the project. D.W. designed and performed the experiment. D.W. analyzed the data; J.O. performed the realistic model simulations; F.M. performed the phase shift analysis. R.Z. and G.G. provided the samples. D.W. wrote the manuscript and drew the figures with the help of J.O., F.M., W.X., Y.Y. and R.W.. All the authors discussed the results.

## Competing interests

The authors declare no competing interests.

## Additional information

**Supplementary information** The online version contains supplementary material available at URL….

**Correspondence** and requests for materials should be addressed to Dongfei Wang (dfwang@bit.edu.cn) and Roland Wiesendanger (wiesendanger@physnet.uni-hamburg.de)

## Figure 1

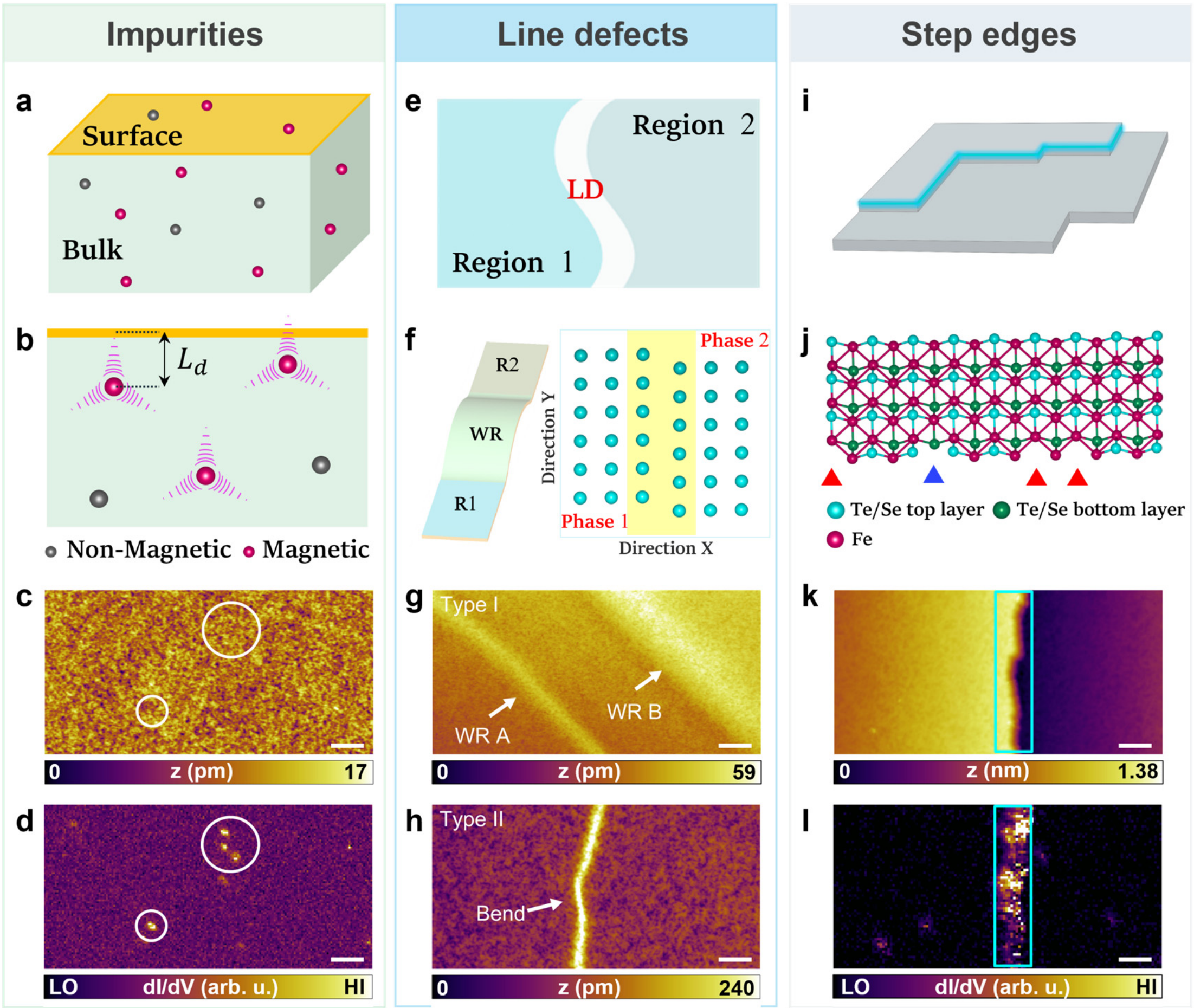

**Fig. 1|Three types of defects on Fe(Te,Se). a**, Illustration of magnetic (purple) and non-magnetic (gray) impurities on the surface as well as inside the bulk. **b**, Magnetic impurities below the surface with in-gap bound states can still be detected by the STM. $L_d$ is the maximum length beyond which the bound states of the bulk impurity can no longer be detected by STM. The symmetry of the bound states do not need to be three-fold. **c**, STM topography image of a surface impurity free area of 80 nm × 160 nm. **d**, dI/dV map at 0 mV of the same region shown in **c**. **e**, Cartoon image of a line defect (LD) on the surface. **f**, Illustration of some typical line defects left, wrinkle (WR) formed due to strain relief; right, grain boundary formed due to in-plane lattice mismatch. **g-h**, STM topography images of the Type I wrinkles (**g**) and the Type II line defect with a bend (**h**) on surface. **i**, Cartoon image of a step edge high-lighted with blue color. **j**, Illustration of the atom lattice model near a step edge. The red and blue triangles point to the positions where atoms are missing during the cleavage. **k**, A vertical step edge on the Fe(Te,Se) surface in a 50 nm × 100 nm area. **l**, dI/dV map at 0 mV of the same region shown in **k**. Scale bar: **c-d** 16 nm; **g** 20 nm; **h**,**k**,**l** 10 nm. Tunneling parameters: **c** $V = -1$ V, $I = -100$ pA; **d** $V_{stab} = -10$ mV, $I_{stab} = -200$ pA, $V_{osc} = 0.03$ mV; **g** $V = -1$ V, $I = -100$ pA; **h** $V = -0.5$ V, $I = -200$ pA; **k** $V = -0.5$ V, $I = -80$ pA; **l** $V_{stab} = 6$ mV, $I_{stab} = 80$ pA, $V_{osc} = 0.03$ mV.

# Figure 2

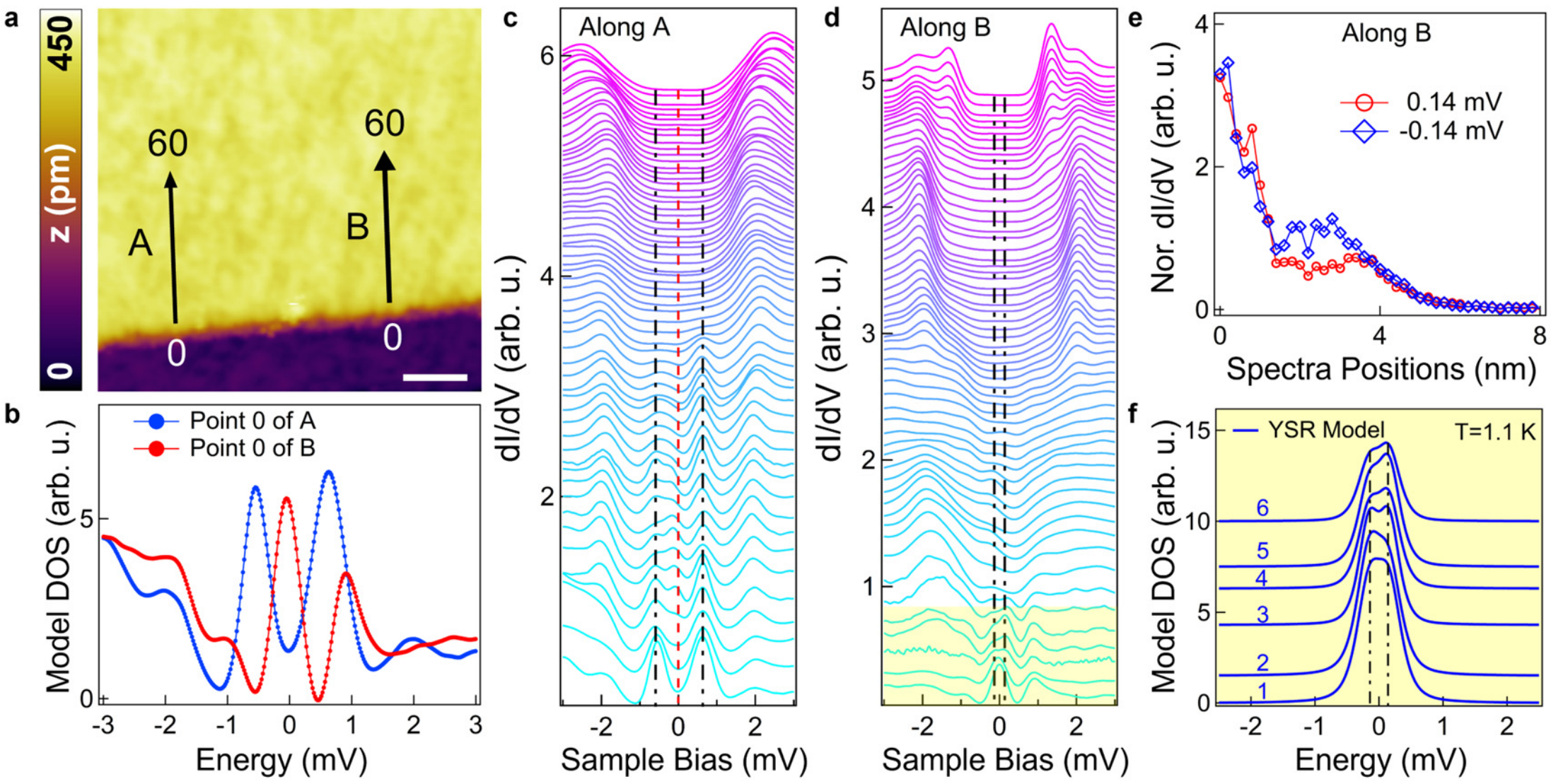

**Fig. 2|Near-zero energy bound state at the step edge of Fe(Te,Se). a**, STM image of a step edge at the bottom part. The two arrows indicate the path along which the dI/dV signals were acquired as shown in **c-d**. **b**, Two dI/dV spectra taken at the step edge (point 0 at the beginning of arrow A and arrow B). **c**, From bottom to top: equally spaced 61 dI/dV spectra taken along the 12 nm long arrow A shown in **a. d**, From bottom to top: equally spaced 60 dI/dV spectra taken along the 12 nm long arrow B shown in **a. e**, Normalized dI/dV signal at 0.14 mV and -0.14 mV extracted from **d** within 8 nm starting from the step edge. **f**, Model plot of the DOS by considering two particle-hole symmetric Lorentzian peaks at $\pm 0.14$ mV with peak heights proportional to the first 6 points shown in **e**. Scale bar: **a** 5 nm. Tunneling parameters: **a** $V = 0.5$ V, $I = 90$ pA; **b-d** $V_{stab} = 5$ mV, $I_{stab} = 100$ pA, $V_{osc} = 0.03$ mV.

# Figure 3

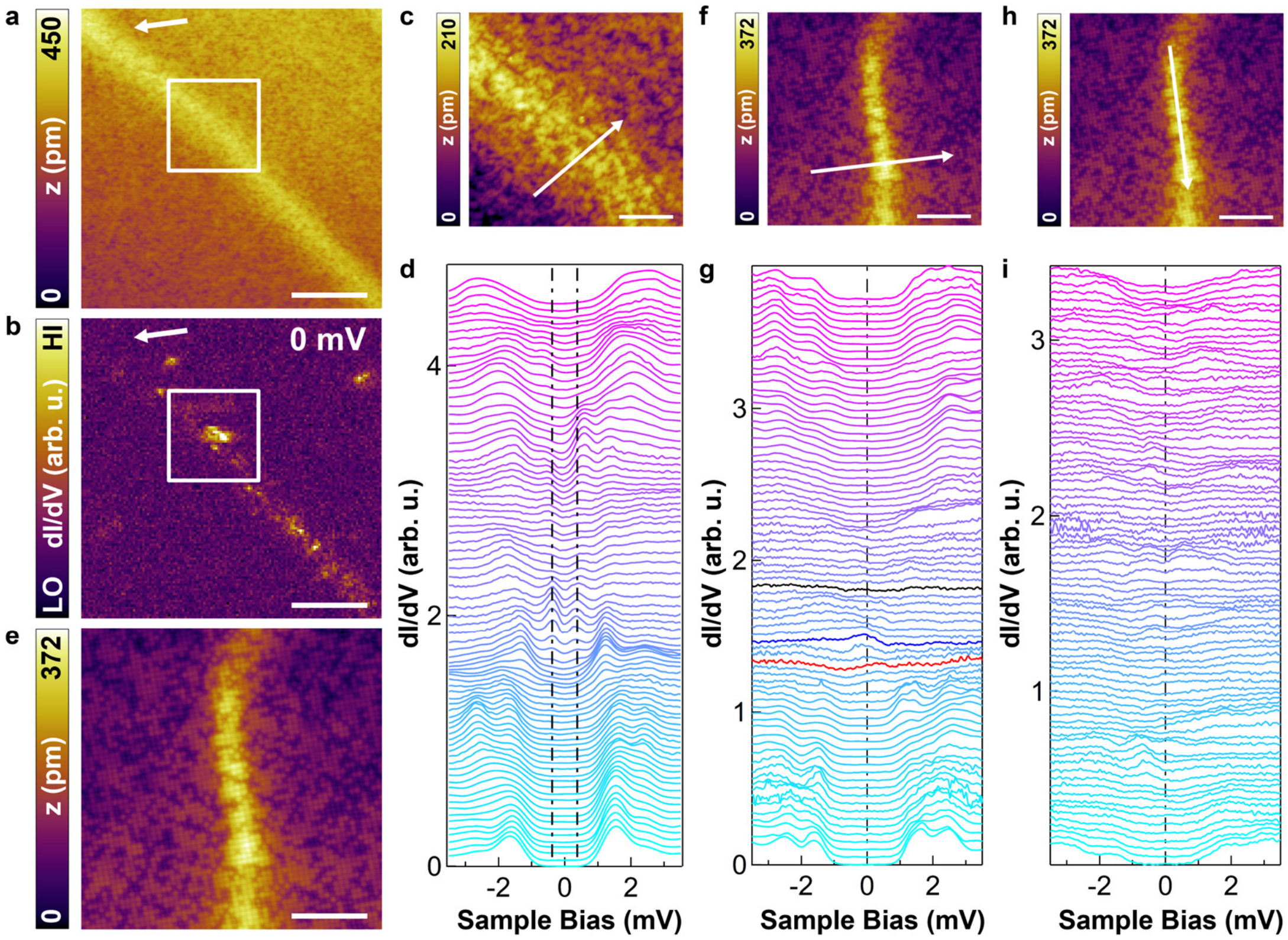

**Fig. 3|In-gap states observed at the Type I wrinkle and Type II line defect bend. a**, STM topography image of a wrinkle on the surface of Fe(Te,Se). **b**, The dI/dV map at 0 mV of the same region shown in **a**. The white box highlights where an enhanced DOS at 0 mV is observed while the white arrow points to the positions where no DOS enhancement at 0 mV is observed at the wrinkle. **c**, Zoom-in STM topography image of the wrinkle region highlighted by the white box in **a**. The white solid arrow indicates where the line spectra are taken in **d**. **d**, From bottom to top: Equally spaced 81 dI/dV spectra taken across the wrinkle along the 16.5 nm long arrow shown in **c**. **e**, STM image of the bend with atomic resolution. **f**, same as **e** with the white arrow indicating where the line spectra are taken in **g**. **g**, From bottom to top: Equally spaced 81 dI/dV spectra along the 13.5 nm long arrow shown in **f** across the bend. Three spectra with typical line shapes are highlighted with red, blue and black colors. **h**, same as **e** with the white arrow indicating where the line spectra are taken in **i**. **i**, From bottom to top: Equally spaced 81 dI/dV spectra along the 13.5 nm long arrow shown in **h** along the bend. Scale bar: **a-b** 25 nm; **c** 7.5 nm; **e**, **f**, **h** 5 nm. Tunneling parameters: **a** $V = -1$ V, $I =$ -100 pA; **b** $V_{stab} = -10$ mV, $I_{stab} = -200$ pA, $V_{osc} = 0.03$ mV. **c** $V = -10$ mV, $I = -100$ pA; **d** $V_{stab} = -10$ mV, $I_{stab} = -200$ pA, $V_{osc} = 0.03$ mV; **e**, **f**, **h** $V = -0.1$ V, $I = -2$ nA; **g**,**i** $V_{stab} = -10$ mV, $I_{stab} = -200$ pA, $V_{osc} = 0.03$ mV.

# Figure 4

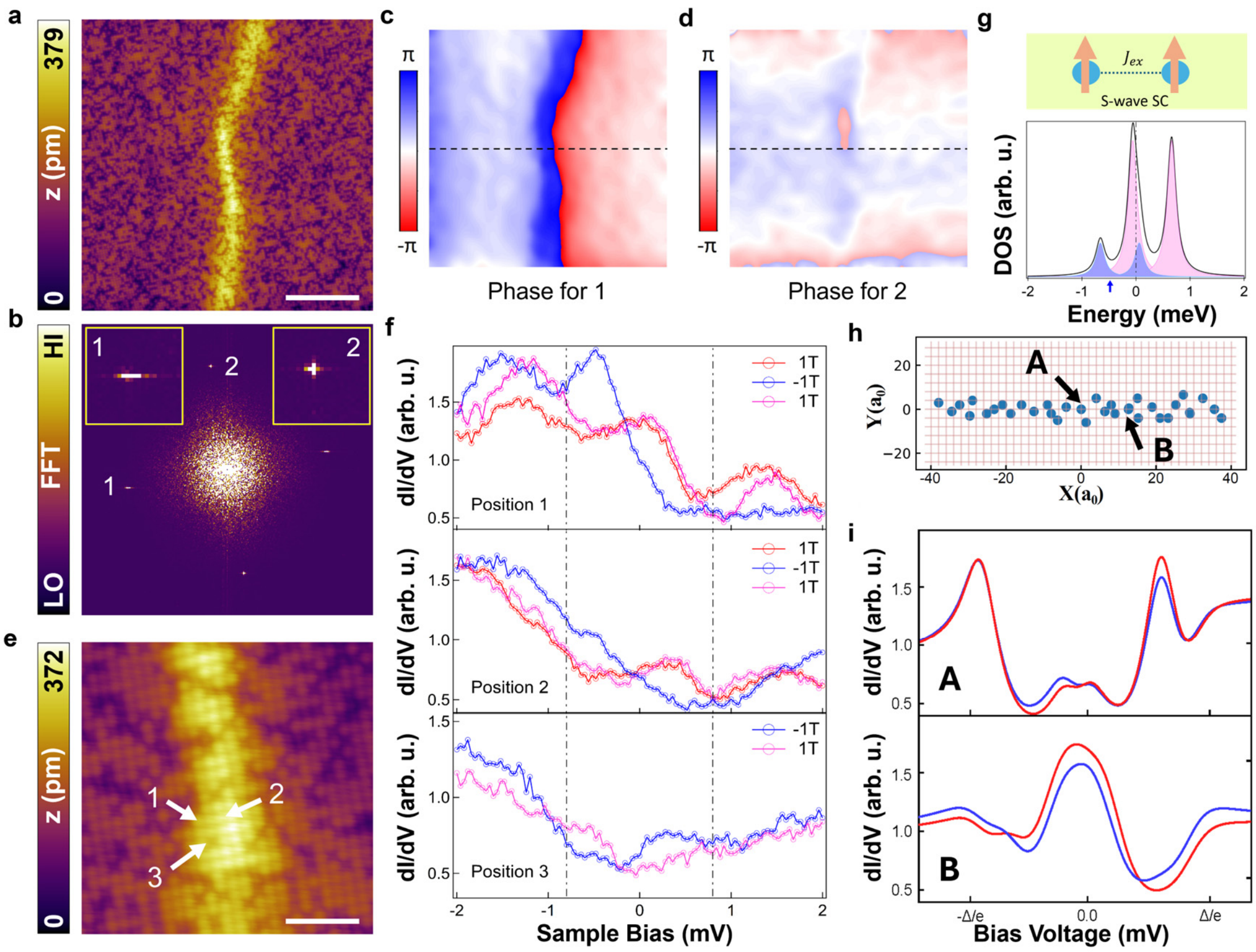

**Fig. 4|Phase shift and spin polarized in-gap state at the Type II line defect bend. a**, STM image of the line defect with atomic resolution. **b**, FFT image of **a** with the reciprocal lattice, the inset highlighted with yellow boxes are zoom-in images of the region near Bragg peak 1 and 2. **c**, phase image of the surface lattice near the bend shown in **a** for the Bragg peak 1 in **b**. **d**, phase image of the surface lattice near the bend shown in **a** for the Bragg peak 2 in **b**. **e**, Zoom-in STM image of the bend, the positions 1-3 are labeled by arrows. **f**, dI/dV spectra near 0 mV at positions 1-3 shown in **e** at 1T, -1T and 1T again. **g**, Top: Schematic image of two identical magnetic impurities in the bulk of a superconductor coupled by ferromagnetic exchange $J_{ex}$. Bottom: The model total DOS of the two ferromagnetically coupled YSR states. Pink stands for the spin-up DOS and blue stands for the spin-down DOS. The blue arrow points to the energy position where the DOS is equal for spin-up and spin-down. **h**, Proposed model for the magnetic impurities below the bend. These impurities are ferromagnetically coupled with different coupling strengths. **i**, A model calculation of spin dependent differential conductance at the two points shown in model **h**. Scale bar: **a** 8 nm; **e** 2.25 nm. Tunneling parameters: **a** $V = 0.1$ V, $I = 2$ nA; **e** $V = -0.1$ V, $I = -2$ nA; **f** $V_{stab} = -10$ mV, $I_{stab} = -300$ pA, $V_{osc} = 0.03$ mV.

# Figure 5

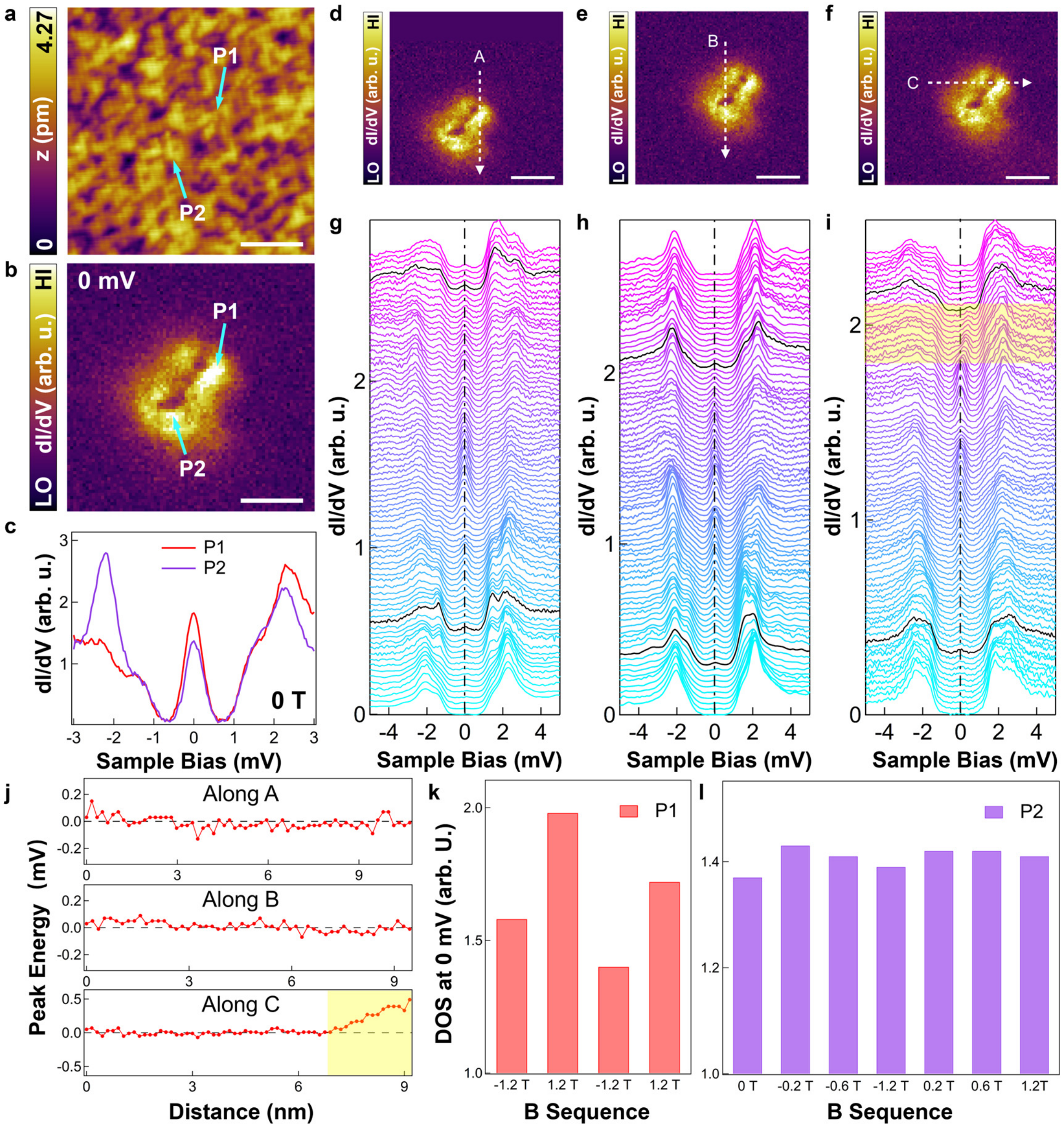

**Fig. 5|Zero-energy bound states for impurities below the surface.** **a**, STM topography of the surface impurity region with atomic resolution. **b**, The dI/dV map at 0 mV of the same region shown in **a**. **c**, The dI/dV taken at point 1 and point 2 indicated by arrows in **a** and **b**. **d-f**, Same as **b** with dashed arrows indicating where the dI/dV in **g-i** are taken. **g-i**, Equally spaced 81 dI/dV spectra along the direction shown in **d-f** for a total length of 12 nm (**g**), 14 nm (**h**) and 12 nm (**i**). The intensity plots are shown in Supplementary Fig. S9 d-f. **j**, The peak energy position within the gap shown in **g-i**, respectively (in between the two black spectra). The yellow high-lighted region shows how the peak position shifts in **i**. **k-l**, Extracted DOS at zero energy at different magnetic fields at point 1 (**k**) and point 2 (**l**). Scale bar: **a, b**, **d**, **e, f** 4 nm. Tunneling parameters: **a** $V$ = -40 mV, $I$ = -200 pA; **b-i** $V_{stab} = -40$ mV, $I_{stab} = -200$ pA, $V_{osc} = 0.03$ mV.

# Supplementary Information: Distinguishing Majorana zero modes from trivial defect states in an iron-based superconductor

Dongfei Wang[1,2,7] ✉, Jon Ortuzar[3,7], Freek Massee[4], Ruidan Zhong[5,6], Genda Gu[5], Wende Xiao[1], Yugui Yao[1], Roland Wiesendanger[2] ✉

[1] School of Physics; Centre for Quantum Physics, Key Laboratory of Advanced Optoelectronic Quantum Architecture and Measurement (MOE); State Key Laboratory of Chips and Systems for Advanced Light Field Display; Beijing Institute of Technology, 100081 Beijing, China
[2] Department of Physics, University of Hamburg, Jungiusstrasse 11, 20355 Hamburg, Germany
[3] CIC nanoGUNE-BRTA, 20018 Donostia-San Sebastián, Spain
[4] Université Paris-Saclay, CNRS, Laboratoire de Physique des Solides, 91405 Orsay, France.
[5] Condensed Matter Physics and Materials Science Department, Brookhaven National Laboratory, Upton, 11973 New York, USA
[6] Tsung-Dao Lee Institute, Shanghai Jiao Tong University, 201210 Shanghai, China
[7] These authors contributed equally: Dongfei Wang, Jon Ortuzar
✉ e-mails: dfwang@bit.edu.cn; wiesendanger@physnet.uni-hamburg.de

- **Supplementary Notes 1-3**
- **Supplementary Figures 1-12**
- **Supplementary Tables 1-2**
- **Supplementary References**

## Supplementary Notes 1-3

## 1. STM line profiles of a step edge and line defects on the Fe(Te,Se) surface

Due to the high quality of the sample, it is not easy to observe a step edge within the scan range of 2000 nm × 2000 nm, Fig. S1a shows such a step edge after searching several surface regions. From the enlarged STM topography image one can clearly see that the step edge is neither straight nor atomically sharp. By plotting the line profile across the step edge, one can see that its height is 1.23 nm which is twice the height of Fe(Te,Se) triple-layers being about 0.6 nm.

On the other hand, we also measured the line profile of Type I wrinkles (WRs, arrows A,B) as well as of a Type II line defect (arrow C,D) as illustrated in Fig. S1c-h. From Fig. S1d and S1e we can see that the peak is slightly asymmetric, which usually is a sign of wrinkle formation due to strain effects. The Type II line defect shown in Fig. S1g and S1h is more symmetric regarding the peak line profile. The height peaks at the line defects are fitted by a Lorentzian function $y_0 + \frac{A}{(x-x_0)^2+B}$ after a linear subtraction $a + bx$ with fitting values shown in Table S1. The corresponding peak height and full width at half maximum (FWHM) are shown in Table S2. Clearly, the line defect bend has the smallest FWHM 2.26 nm and biggest peak height 0.19 nm which makes it different from the WRs.

## 2. Calibration of spin-polarized STM tip

The magnetic tip used in this experiment is a bulk Cr tip prepared by a method described in previous literature [1]. A high temperature flash above 1500 K of the tip in ultrahigh vacuum was applied to remove the surface oxide layer. We used the double-layer Fe film on W(110) with out-of-plane magnetization to check the tip's out-of-plane spin orientation [2]. As bulk Cr is an antiferromagnetic material, it is not possible to switch the tip magnetization below 3 T, which is the maximum field available in our instrument. Therefore, we crashed the tip into the Fe double-layer for picking up some Fe cluster.

In this way, we were able to switch the tip's spin direction in a magnetic field lower than 1 T.

For the spin-polarized STS measurement shown in Fig. 4, the magnetic field response of the Cr tip 1# is shown in Fig. S7. To make our spin-polarized measurement reliable, we usually chose stable magnetic tips obtained after long-time scanning, for example 6 hours, meaning that the tip's spin orientation did not change, as seen in Fig. S7a and S7b. We ramped our magnetic field in both positive and negative directions perpendicular to the surface to check the critical field for which the tip's spin direction changes. From Fig. S7c, one can see that at higher magnetic field of 0.6 T, the Fe double-layer has a magnetization direction anti-parallel to the tip. By changing the magnetic field to a negative value (-0.1 T and -0.2 T), only the Fe double-layer at the right bottom switches its magnetization parallel to the tip's magnetization direction. Further decreasing the magnetic field to -0.3 T, we can see that the tip changes its magnetization direction as the Fe double-layer labeled by the white arrows reversed its contrast, as seen in Fig. S7f. In the meantime, the Fe double-layer domains labeled by the yellow dashed arrow also switched their magnetization as their contrast didn't change. Further decreasing the magnetic field polarized all the Fe double-layer patches into the direction anti-parallel to the tip's magnetization direction, as shown in Fig. S7g and S7h. By calculating the spin asymmetry of the tunnel junction using the equation $P = (S_P - S_{AP})/(S_P + S_{AP})$, where $S_P$ and $S_{AP}$ are the averaged dI/dV signals for bright and dark areas in Fig. S7f, we get $P = 12\%$.

Similarly, for the spin-polarized STS measurement shown in Fig. 5, the magnetic field response of the Cr tip 2# is shown in Fig. S8. From Fig. S8b-c and S8e-f one can clearly see that the tip changes its magnetization direction as the Fe double-layer domains labeled by the white arrows switch their contrast. At higher magnetic field of +1 T and -1 T, the Fe thin film becomes fully magnetized into a direction anti-parallel to the tip spin direction as seen in Fig. S8g and S8h. Similar tunnel junction's spin asymmetry analysis of Fig. S8e yields a value of 21.2%. Overall, the Fe cluster decorated Cr tip is very stable and changes its magnetization direction in an external magnetic field higher than 0.4 T.

## 3. Theoretical model of the YSR states on Fe(Te,Se).

### 1) Toy model

The density of states of the first 6 points near zero energy shown in Fig. 2f are simulated by the summation of two Cauchy-Lorentz distributions defined by $amp/b\pi(1+(-a+x)^\wedge 2/b^\wedge 2\,)$. Here, $a$ is the peak energy for which we chose $\pm 0.16\,mV$. The peak width is modeled by b and for the best fit, b is 0.08 for the spectra 1-2 and 0.05 for spectra 3-6. The parameter b reflects the coupling strength between the magnetic impurity and the substrate. Its two values for the best fit indicate that there are two impurities involved in the bound state formation. The $amp$ is the amplitude of the Lorentz function and we chose a value that is proportional to the peak height value extracted from Fig. 2f at $\pm 0.16\,mV$, as presented in Fig. S2b. At 0 K, without considering temperature broadening effects, the two peaks can clearly be identified even though their energy positions are very close to zero, as seen in Fig. S2a. However, we can add the temperature-effect by convoluting the model DOS $\rho_s$ with the Fermi-Dirac distribution at 1 K using the following equation:

$$\frac{dI}{dV} = Const.\times \frac{d}{dV}\int_{-\infty}^{+\infty} \rho_s(\epsilon + eV) \times [f(\varepsilon) - f(\epsilon + eV)]d\epsilon$$

Here, $f(\varepsilon)$ is the Fermi-Dirac distribution function and $Const. = 1$. We find that the peaks at $\pm 0.16\,mV$ merge into one peak at the first point and look like they are splitting at points 2 to 6, as shown in Fig. 2d.

### 2) Realistic model

The Yu-Shiba-Rusinov (YSR) approximation for conventional s-wave superconductors [3-5] can be naturally extended to more complex systems, such as s±-wave superconductors, by considering the coupling of a magnetic impurity to multiple bands [6]. Since the precise nature of this coupling for the system simulated here is not known, it lies beyond the scope of our analysis. Furthermore, variations in the coupling strength would only lead to minor modifications in the formation of in-gap states.

The aim of our simulations is to demonstrate that the in-gap density of states (DOS) observed in Fe(Te,Se) can be explained by the presence of YSR states. For this purpose, we simplify the model to local magnetic impurities interacting with an s-wave superconductor. While our model does not attempt to reproduce the detailed gap structure of Fe(Te,Se), our focus is solely on the in-gap dynamics.

We model the system as an ensemble of local impurities:

$$\hat{V} = \sum_{i}^{N} \widehat{\psi_\sigma^\dagger}(\vec{r_i})(J_i\vec{S_i}\overrightarrow{s_{\sigma\sigma'}} + U\delta_{\sigma\sigma'})\widehat{\psi_\sigma}(\vec{r_i})$$

where $\widehat{\psi_\sigma^\dagger}(\vec{r_i})\left(\widehat{\psi_\sigma}(\vec{r_i})\right)$ are the creation (annihilation) operators at positions $\vec{r_i}$, while $J_i$, U, and $\vec{S_i}$ are the exchange coupling, scalar scattering potential and spin direction (classical spin) of the i$^{\text{th}}$ impurity, and $\overrightarrow{s_{\sigma\sigma'}}$ is the substrate spin operator. Moreover, we describe the sample as a 2D superconductor

$$\widehat{H_0} = \int d\vec{r}\, \widehat{\psi_\sigma^\dagger}(\vec{r})\, \xi\widehat{\psi_\sigma}(\vec{r}) + \Delta\widehat{\psi_\sigma^\dagger}(\vec{r})\widehat{\psi_{-\sigma}^\dagger}(\vec{r}) + h.c.$$

where $\Delta$ is the superconducting gap. The dynamics of the in-gap states are qualitatively equal for 2D or 3D superconductors. The decay of the YSR state extension can change with dimensionality [7], but the precise extension of these states is not important for our analysis.

In the following we will prove three concepts: a) A dispersing zero bias peak can be explained by two interacting impurities. b) An ensemble of magnetic impurities can describe the enhanced DOS at zero bias on the bend of Type II line defect. c) The combination of thermal broadening and particle-hole asymmetry can give a finite spin polarization of a zero bias YSR state.

**Simulating a splitting zero bias peak with two YSR peaks close to zero**

Figure 2b in the main text shows a zero-bias peak at a step edge. As consecutive spectra are taken along a line away from the impurity, the zero-bias peak shifts in energy. This excludes the existence of a Majorana zero mode. This kind of topological state appears only at zero energy. Any shift away from it resembles more a YSR state. We propose that the in-gap states measured at the step edges arise from local magnetic moments at the edge, as sketched in Fig. S3a. Fig. S3b presents the simulation of the dI/dV measured along a line away from one of the impurities.

We note that the peak that appears centered at zero bias moves slightly to positive energies close to the impurity and then to negative energies further away from it. This is due to the extension of the YSR state of the second impurity. Due to the oscillatory behavior of the spatial extension of YSR states, the overlap between two distinct impurities is position dependent, and hence, the position of the peak in the DOS can shift. This property can be used to engineer any kind of shift of the YSR state. The effect in this case is small, but sufficient to disprove the existence of a Majorana mode.

**Simulation of the Type II line defect.**

Simulating the precise in-gap DOS at the bend is not possible with limited information available from STM experiments. Critical parameters such as the exact positions of magnetic impurities and their exchange coupling strengths with the superconducting substrate remain unknown.

Nevertheless, the methodology introduced in Ref. [7,8] provides a framework for calculating the impact of an ensemble of magnetic impurities on a superconductor. Using the open-source code available in Ref. [9], we simulate the effect of an impurity ensemble arranged to qualitatively mimic the spatial structure of the line defect bend, as shown in Fig. S10a.

This simulation serves as a sanity check, qualitatively demonstrating that magnetic impurities offer the simplest and most plausible explanation for the observed enhancement in the zero-bias DOS.

Figure S10a shows a random distribution of magnetic impurities used in the simulation. The distribution resembles that of the bend found in the experiments. The exchange coupling of each impurity is a random distribution around $J_C$, i.e,

$$V_i = (J_C + \delta J)\vec{S_i}\overrightarrow{s_{\sigma\sigma'}} + \delta U \delta_{\sigma\sigma'}$$

where $E_{YSR}(J = J_C) = 0$, $\delta J \in [-0.1, 0.1]$, and $\delta U \in [-0.4, 0.4]$. This choice reflects a random distribution of exchange couplings around the quantum phase transition (QPT) point, where the YSR states cross zero energy. Additionally, we include a finite scalar potential term $\delta U$ to account for the particle-hole asymmetry observed experimentally.

Figure S10b displays the resulting spatial map of the zero-energy DOS, which corresponds directly to the zero-bias conductance measured in scanning tunneling spectroscopy (STS). Vertical and horizontal line profiles, taken along the red crosses in Fig. S10a and shown in Fig. S10c, further illustrate the local enhancement of the zero-bias differential tunneling conductance, obtained by convoluting the DOS with a Fermi-Dirac distribution, across the simulated impurities below the bend.

This calculation demonstrates that a finite-sized spatial enhancement of the zero-bias conductance—similar to the experimental observations—can naturally arise from a disordered distribution of magnetic impurities near the QPT. For a quantitative, system-specific simulation, however, the exact spatial locations of the magnetic impurity cores and their coupling strengths would be required—information that remains inaccessible in STM experiments.

**Zero-bias spin polarization**

One of the hallmark signatures of a Majorana bound state is the appearance of a zero-bias peak in spectroscopic measurements. However, other physical phenomena can also give rise to zero-bias conductance features, such as the Josephson effect [10], Andreev bound states [10], or Yu-Shiba-Rusinov (YSR) states [3-5].

In the present experiment, the Josephson effect between tip and sample can be excluded, as a non-superconducting tip is employed. Moreover, Andreev reflections at the tip–sample interface typically occur only at low resistances. Since all measurements here are performed in the tunneling limit (R>1 MΩ), Andreev reflections can also be disregarded.

As discussed earlier, the simplest explanation for a finite in-gap conductance is the presence of magnetic impurities, which lead to the formation of YSR states. At the quantum phase transition (QPT), the positive- and negative-energy YSR peaks merge, producing a zero-bias peak composed of two degenerate states. In this case, the spin polarization at zero bias — in contrast to a Majorana bound state — is expected to vanish: the contributions from the two YSR peaks cancel, resulting in no measurable spin-polarized conductance.

On the other hand, if more than one impurity is present in the sample, a zero-bias peak can also arise due to the hybridization of YSR wavefunctions, forming bonding and antibonding YSR states. Such a peak does not correspond to an exact overlap of the positive and negative energy peaks and can therefore exhibit a finite spin polarization. In conclusion, the observation of spin polarization at zero bias does not constitute definitive evidence for the presence of a Majorana bound state. Instead, an ensemble of magnetic impurities could account for the same behavior. As an illustrative example, Fig. S11 shows the calculated spin-polarized dI/dV for a system with two impurities (see Fig. S11a). At the midpoint between the two impurities (Fig. S11b), the spin polarization at zero bias vanishes, while at the impurity sites (Fig. S11c), a pronounced difference between spin-up and spin-down dI/dV is observed at zero bias.

# Supplementary Figures 1-11

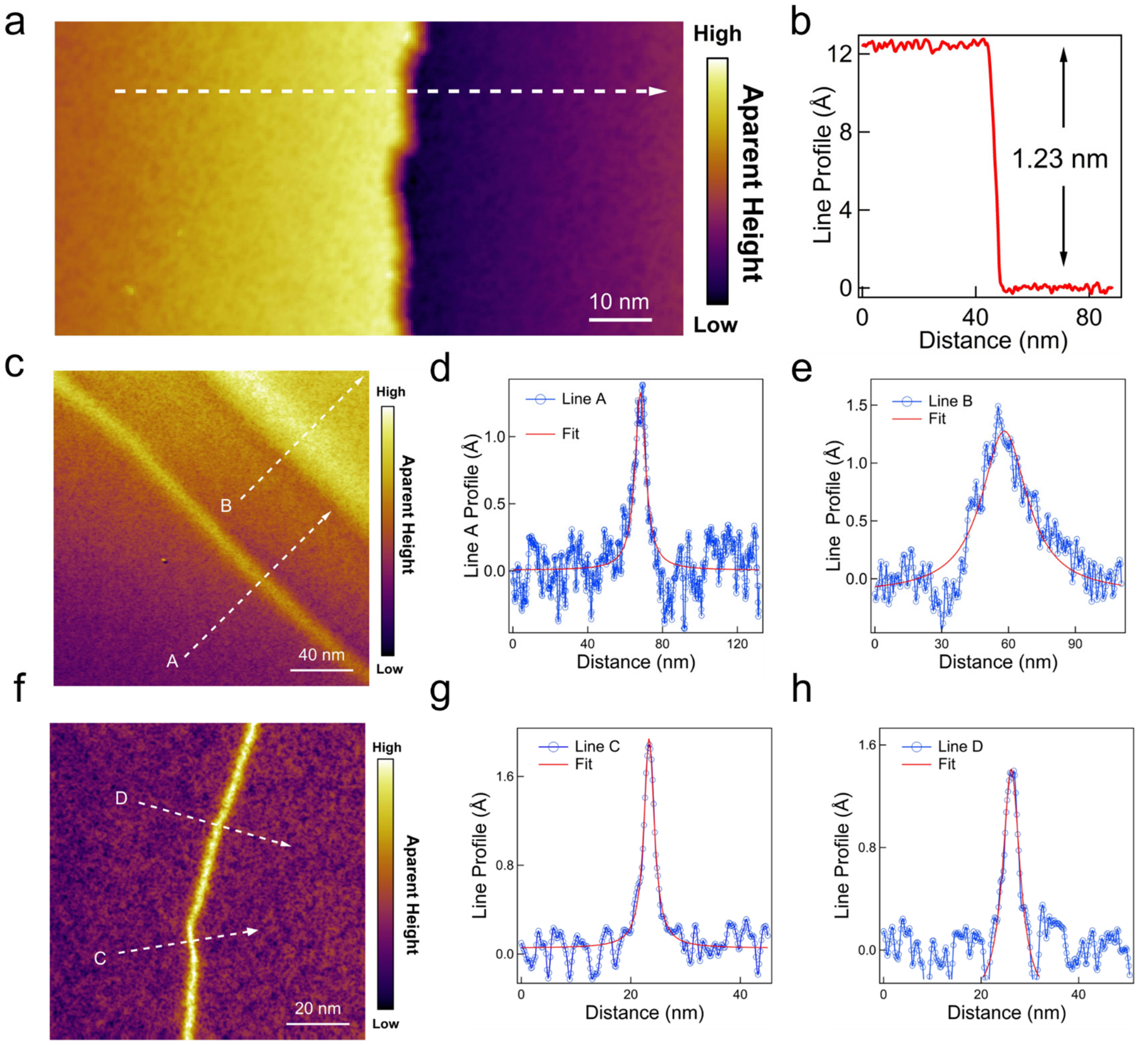

**Fig. S1|Line-profiles for step edge and line defects on Fe(Te,Se). a**, Same STM topography image shown in Fig. 1k with dashed arrow indicating where the line-profile is taken in **b**. **b**, Line-profile of the step edge along the path shown in **a**. **c**-**e**, STM topography image of the same region shown in Fig. 1g (**c**) and the line-profiles along the path A (**d**) and path B (**e**) shown in **c** correspondingly. The red curves are the Lorentzian fits. **f**-**h**, STM topography image of the same region shown in Fig. 1h (**f**) and the line-profiles along the path C (**g**) and path D (**h**) shown in **f** correspondingly. The red curves are the Lorentzian fits. Tunneling parameters: **a** $V = -0.5$ V, $I = -80$ pA; **c** $V = -1$ V, $I = -100$ pA; **f** $V = -0.5$ V, $I = -200$ pA.

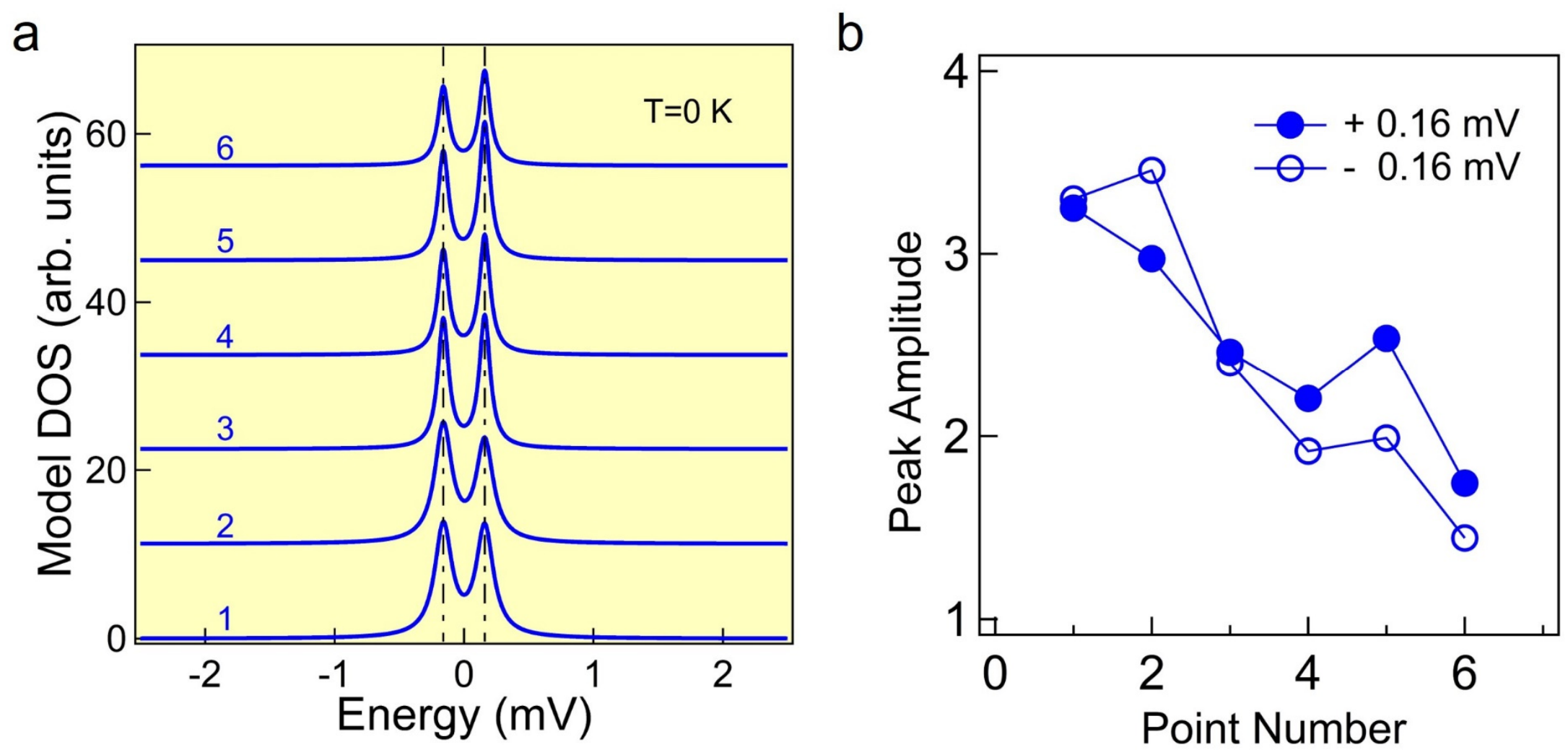

**Fig. S2|Toy model fit of the two closely packed YSR bound states. a**, YSR bound states fit near zero energy using the summation of two Cauchy-Lorentz distributions with particle-hole symmetry at 0 K. **b**, the peak amplitudes from **a** at $\pm 0.16$ mV. The values are proportional to the DOS values extracted at the same energy from Fig. 2d for the first 6 points.

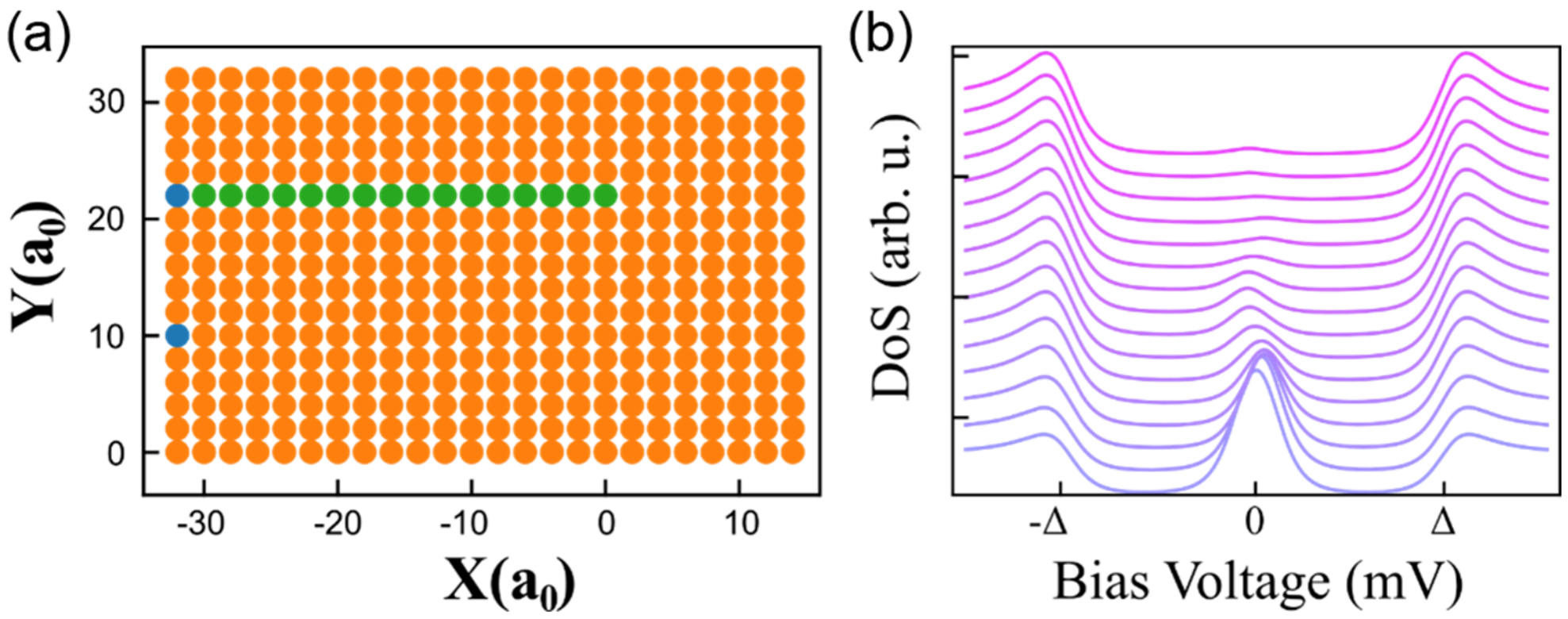

**Fig. S3|Model fit of the two closely packed YSR bound states. a**, Spatial distribution of the magnetic impurities on top of the material. Blue dots represent the magnetic impurities, the orange dots correspond to the lattice of the material, and the green dots are the positions where the calculation of the DOS in **b** is done. **b**, Simulated DOS of the system for the green points in panel **a**.

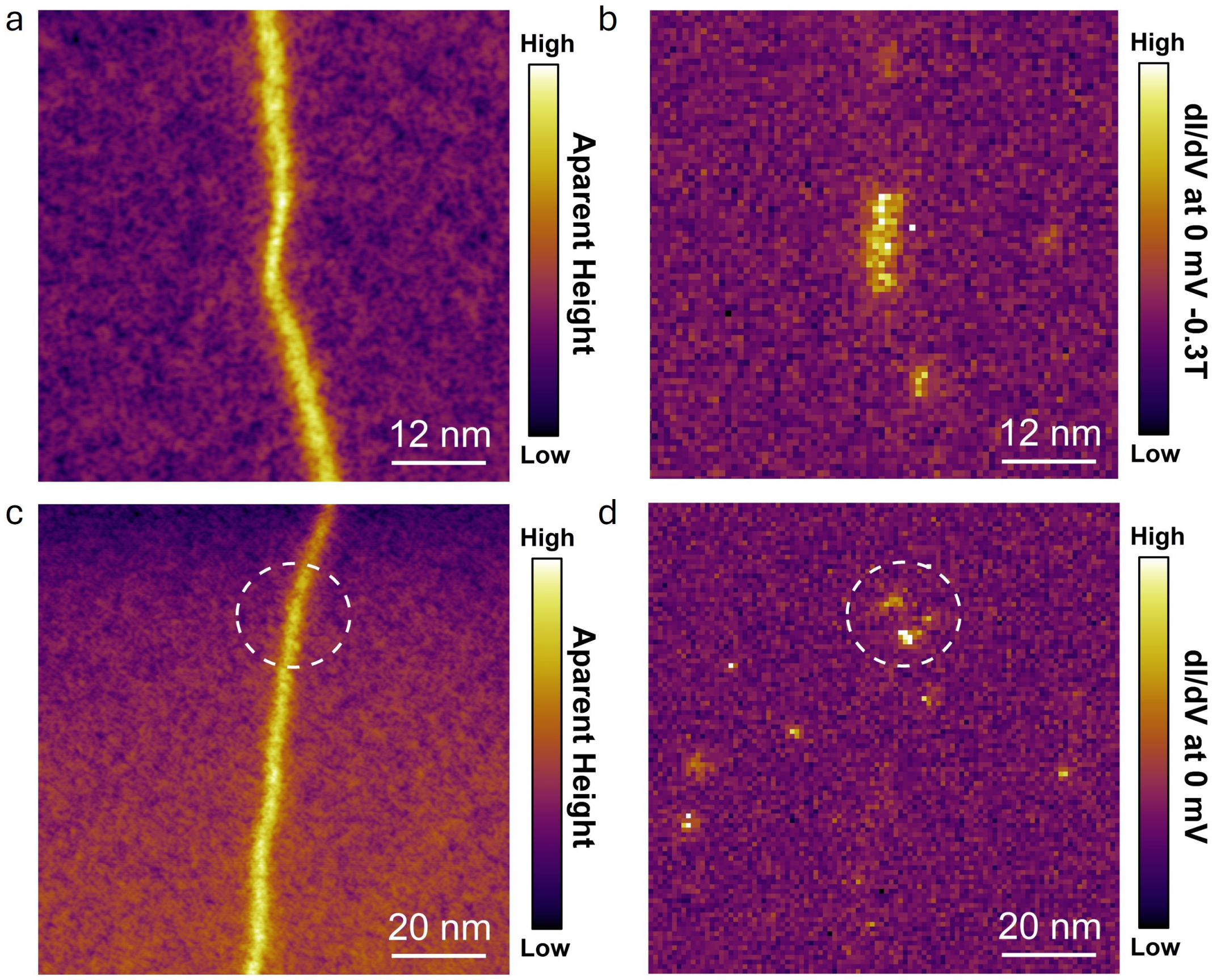

**Fig. S4|Zero-energy DOS at Type II line defects. a-b**, STM topography image of a type II line defect with a bend (**a**) and dI/dV map at 0 mV of the same region (**b**). **c-d**, STM topography image of another Type II line defect (**c**) and dI/dV map at 0 mV of the same region (**d**). Tunneling parameters: **a** $V = -0.5$ V, $I = -200$ pA; **b** $V_{stab} = -10$ mV, $I_{stab} = -200$ pA, $V_{osc} = 0.03$ mV; **c** $V = -0.1$ V, $I = -200$ pA; **d** $V_{stab} = -10$ mV, $I_{stab} = -200$ pA, $V_{osc} = 0.03$ mV.

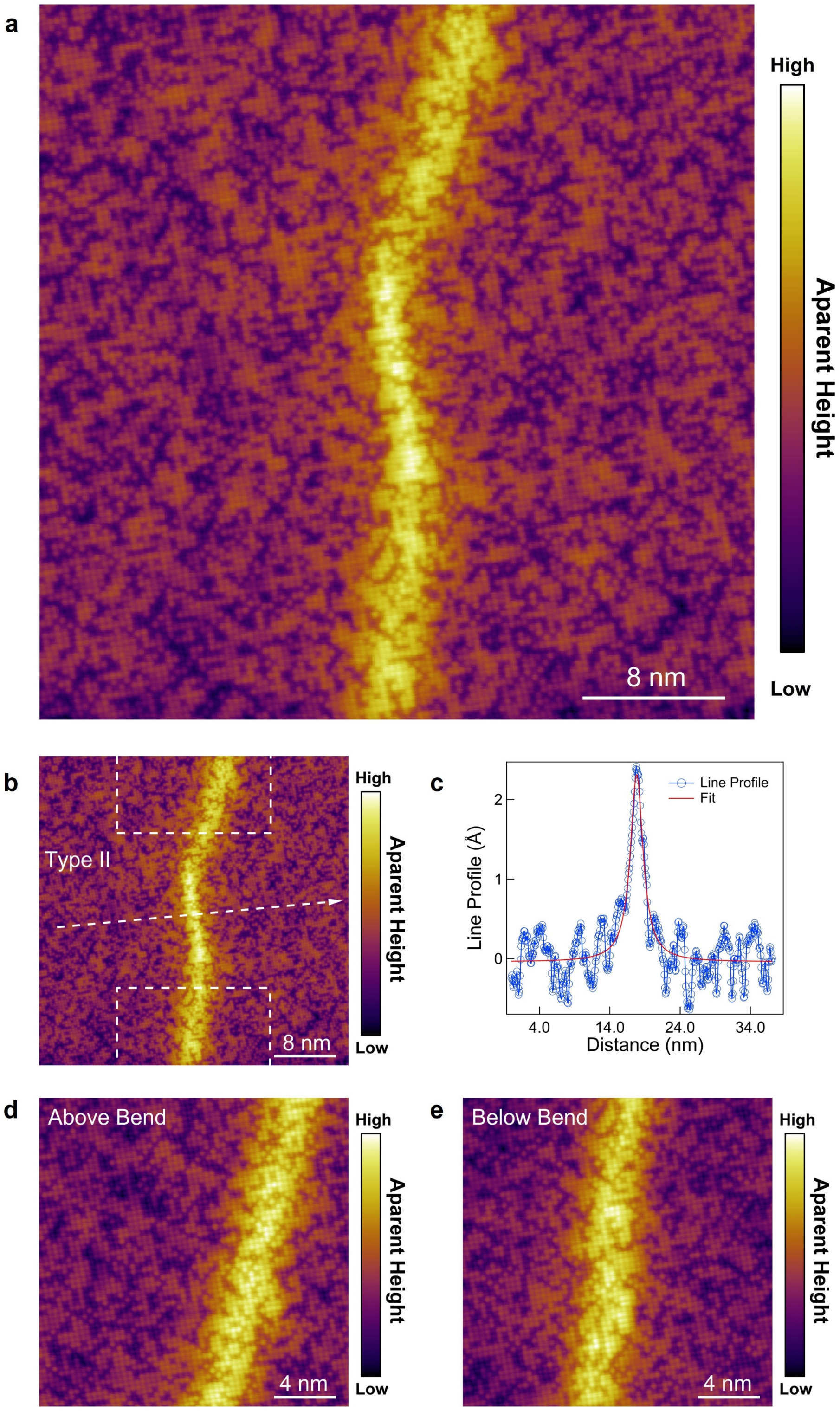

**Fig. S5|STM topography image and line-profile at the Type II line defect with atomic resolution. a**, STM topography near bend with atomic resolution. **b**, same with **a**, the dashed arrow indicates where the line-profile in **c** is taken. **c**, line-profile across the bend along the path shown in **b**. **d-e**, STM topography images of the line defects above (**d**) and below (**e**) the bend for the regions indicated by the dashed boxes in **b**. Tunneling parameters: **a-b** $V = 0.1$ V, $I = 2$ nA; **d-e** $V = -0.1$ V, $I = -2$ nA.

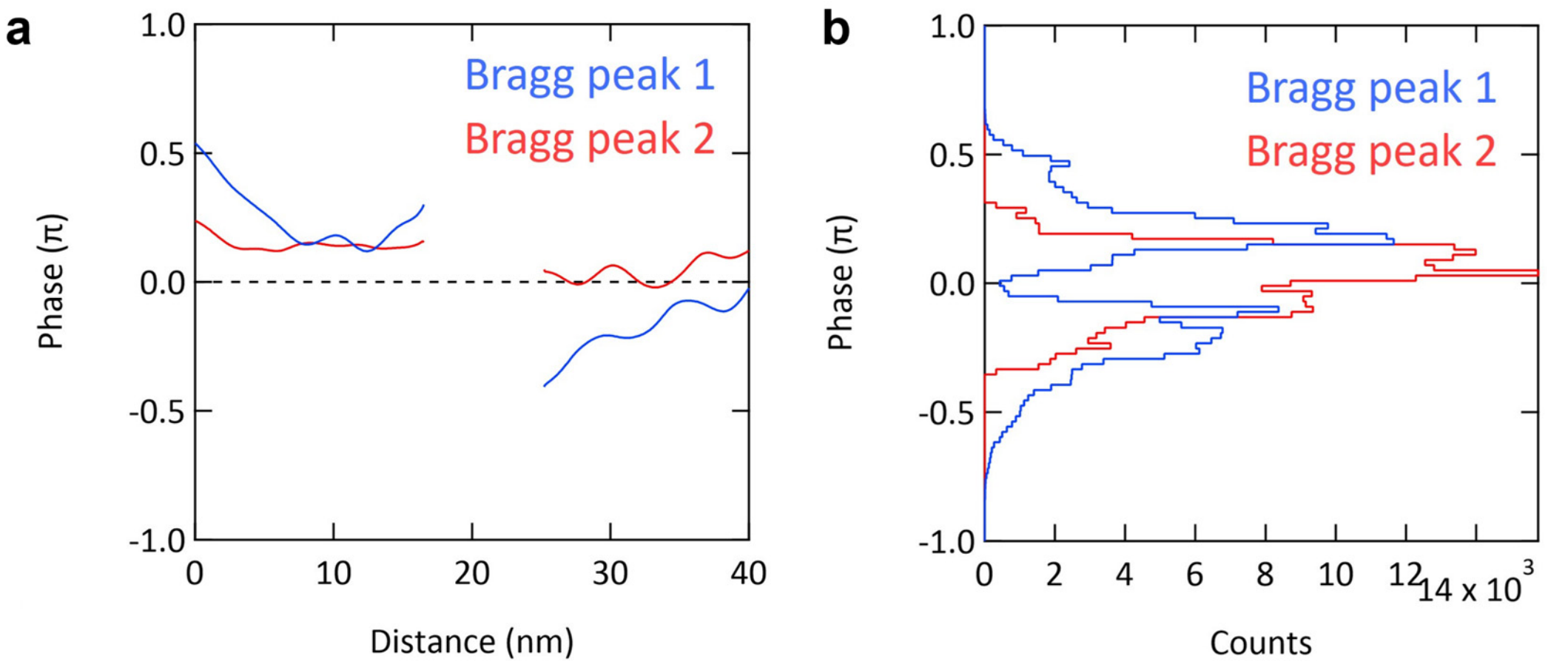

**Fig. S6|Phase shift across the line defect bend. a**, Extracted phase values along the dashed lines shown in Fig. 4c and Fig. 4d from left to the right. **b**, Histogram of the full phase images in main text Fig. 4c,d. The phase difference between either side of the line defect can accurately be extracted from the peaks in the histogram, giving a total phase difference on the order of $\pi / 2$.

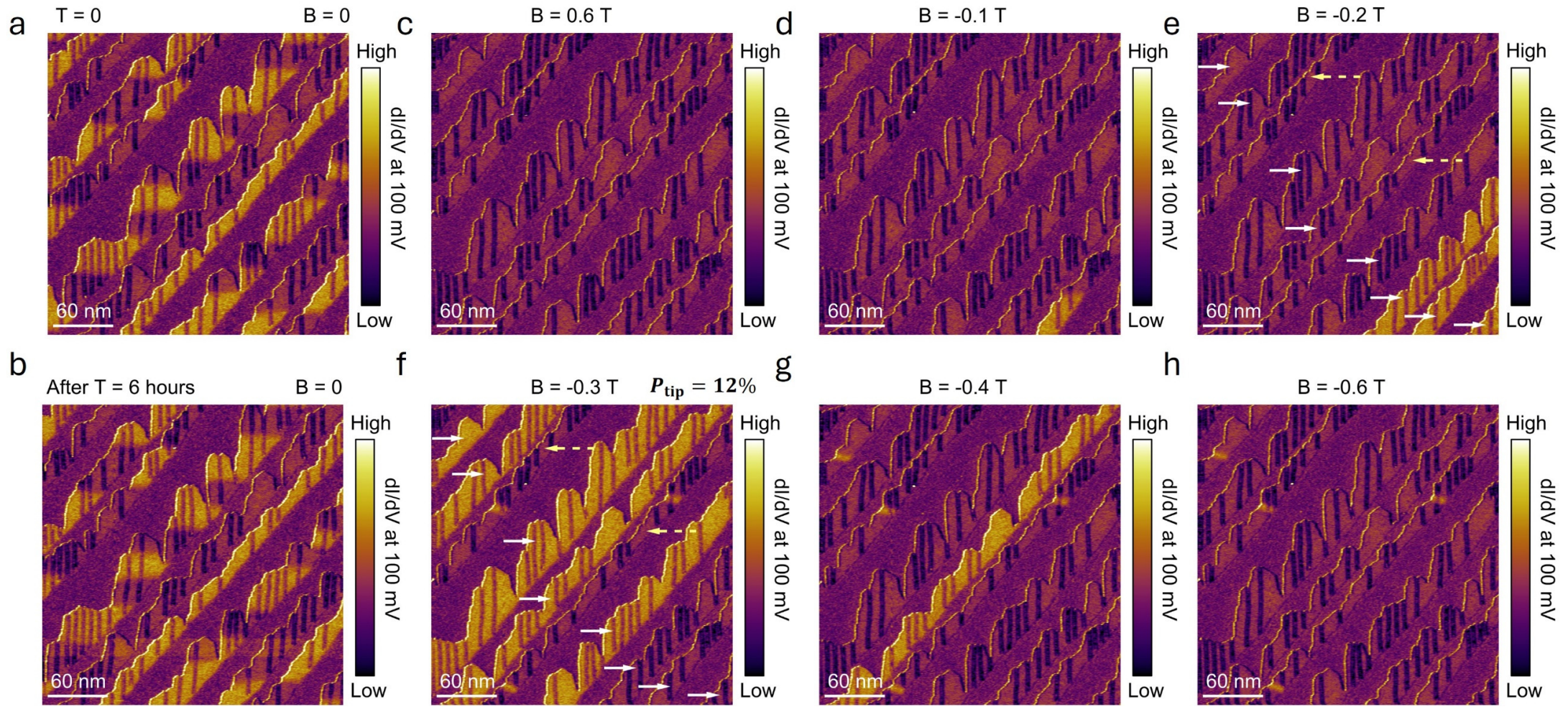

**Fig. S7|Spin-polarized tip #1 calibration at different magnetic fields. a-b**, Spin-resolved dI/dV maps at 100 mV at the beginning (T=0) and after 6 hours (T= 6 hours). **c-h**, Spin-resolved dI/dV maps at 100 mV of the same region at 0.6 T (**c**), -0.1 T (**d**), -0.2 T (**e**), -0.3 T (**f**), -0.4T (**g**) and -0.6 T(**h**), correspondingly. An average of 12% spin asymmetry of the tunnel junction is deduced by analyzing the data shown in **f**. Tunneling parameters: **a-h** $V =$ 100 mV, $I =$100 pA, $V_{osc} =$ 13 mV.

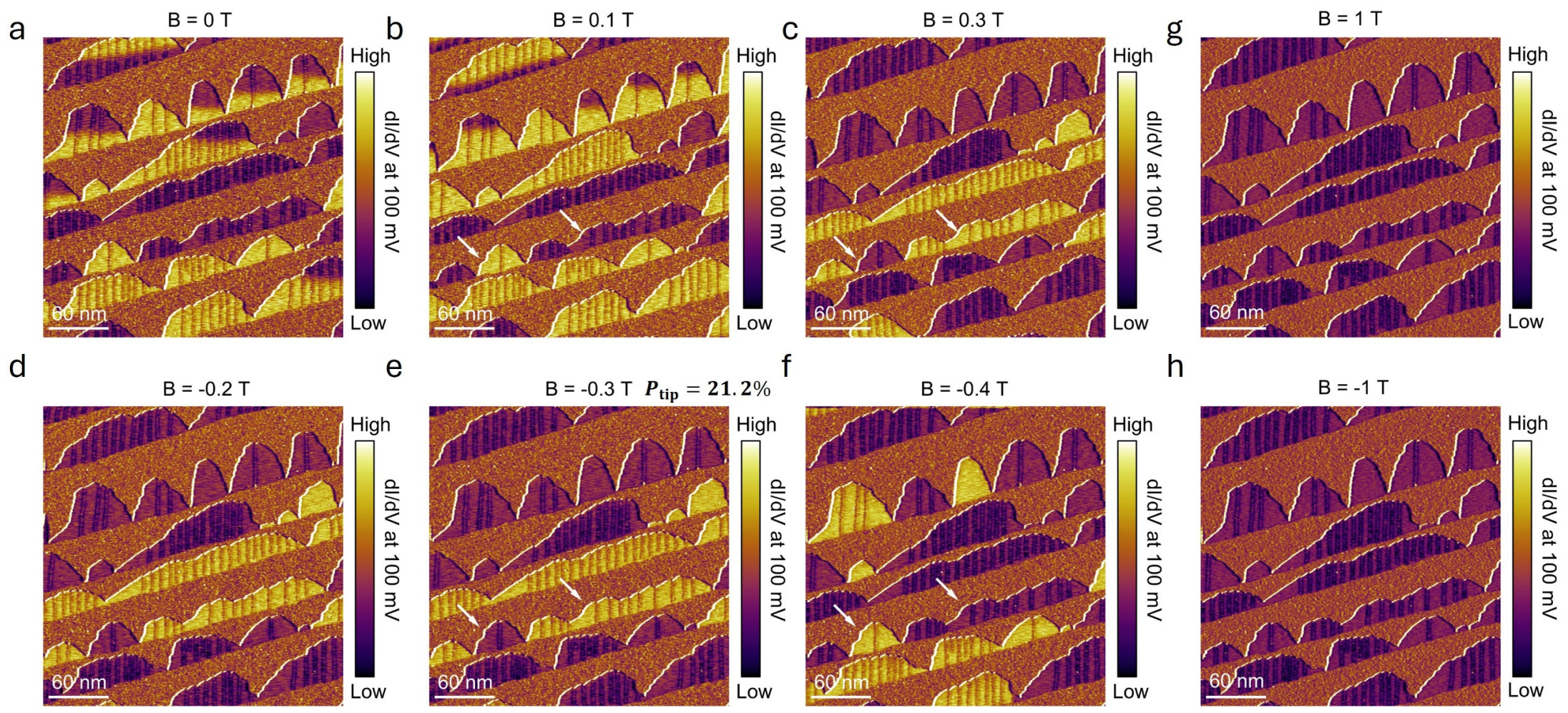

**Fig. S8|Spin-polarized tip #2 calibration at different magnetic fields. a-h**, Spin-resolved dI/dV maps at 100 mV of the same region at 0 T (**a**), 0.1 T (**b**), 0.3 T (**c**), -0.2 T (**d**), -0.3 T (**e**), -0.4 T (**f**), 1T (**g**) and -1 T(**h**) correspondingly. An average of 21.2% spin asymmetry of the tunnel junction is deduced by analyzing the data shown in **e**. Tunneling parameters: **a-h** $V =$ 100 mV, $I =$100 pA, $V_{osc} =$ 13 mV.

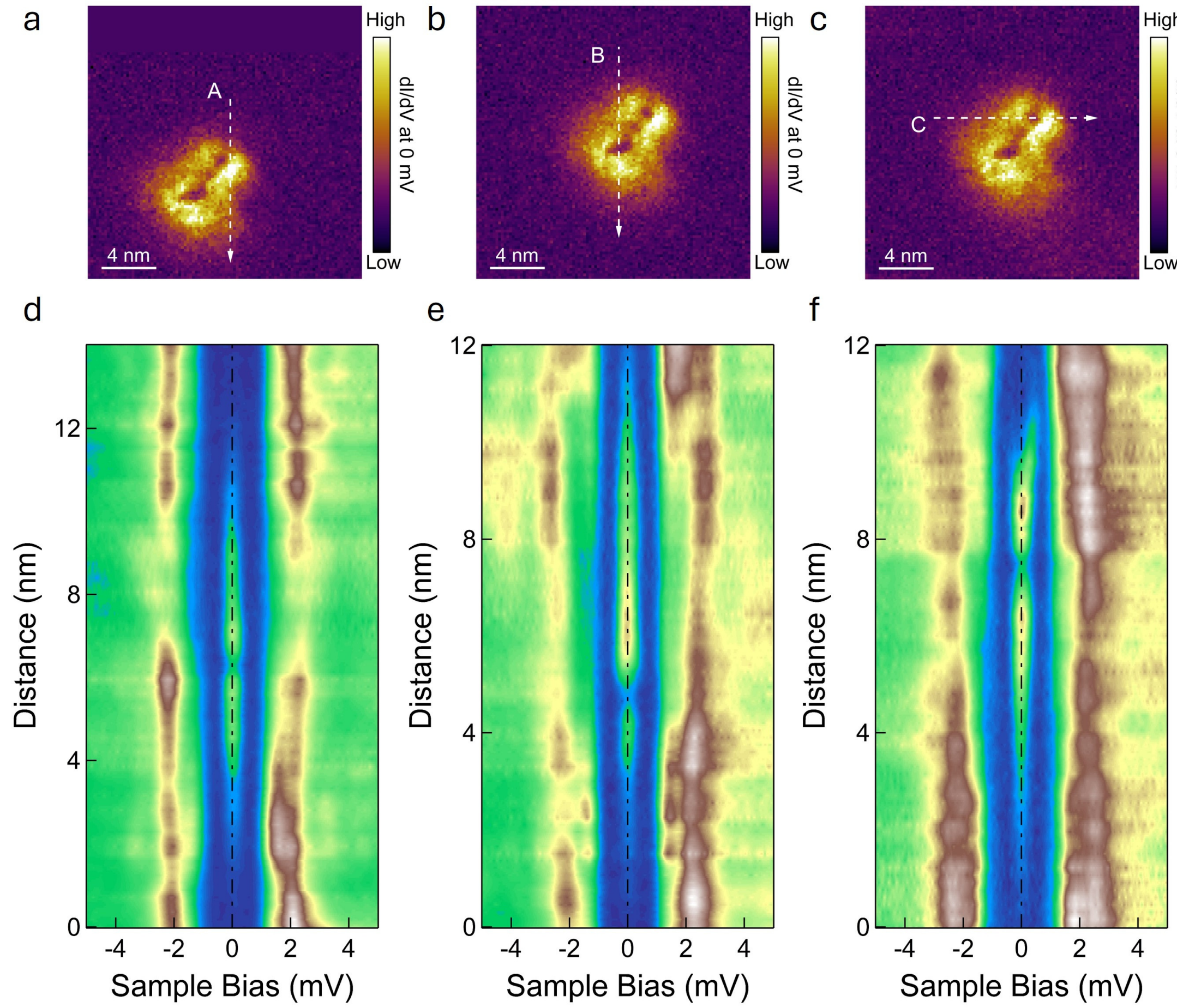

**Fig. S9|dI/dV intensity plots near the impurity below the surface.** **a**-**c**, STM-based dI/dV maps at 0 mV as in Fig. 5d-f. The white dashed arrows indicate the paths along which the intensity plots in **d**-**f** are taken. **d**-**f**, The intensity plots of the dI/dV spectra shown in Fig. 5g-I, correspondingly. Tunneling parameters: **a**-**f** $V_{stab}=-40$ mV, $I_{stab}=-200$ pA, $V_{osc}=0.03$ mV.

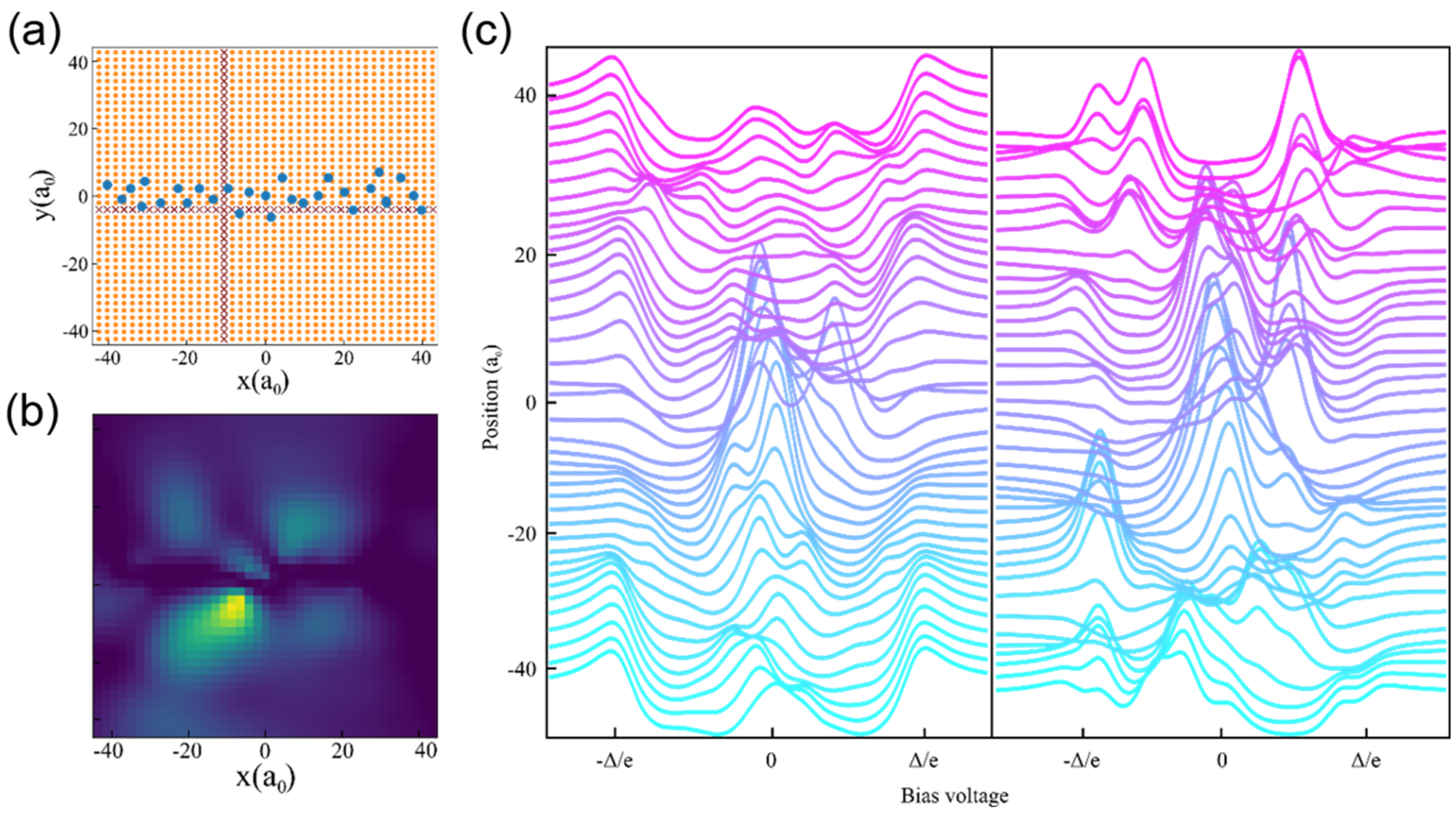

**Fig. S10|Model fitting of the line defect.** **a**, Distribution of the magnetic impurities on the surface of a 2D s-wave superconductor. **b**, Zero energy DOS map. **c**, Line profiles of the DOS, following the vertical and horizontal red crosses in panel **a**, respectively.

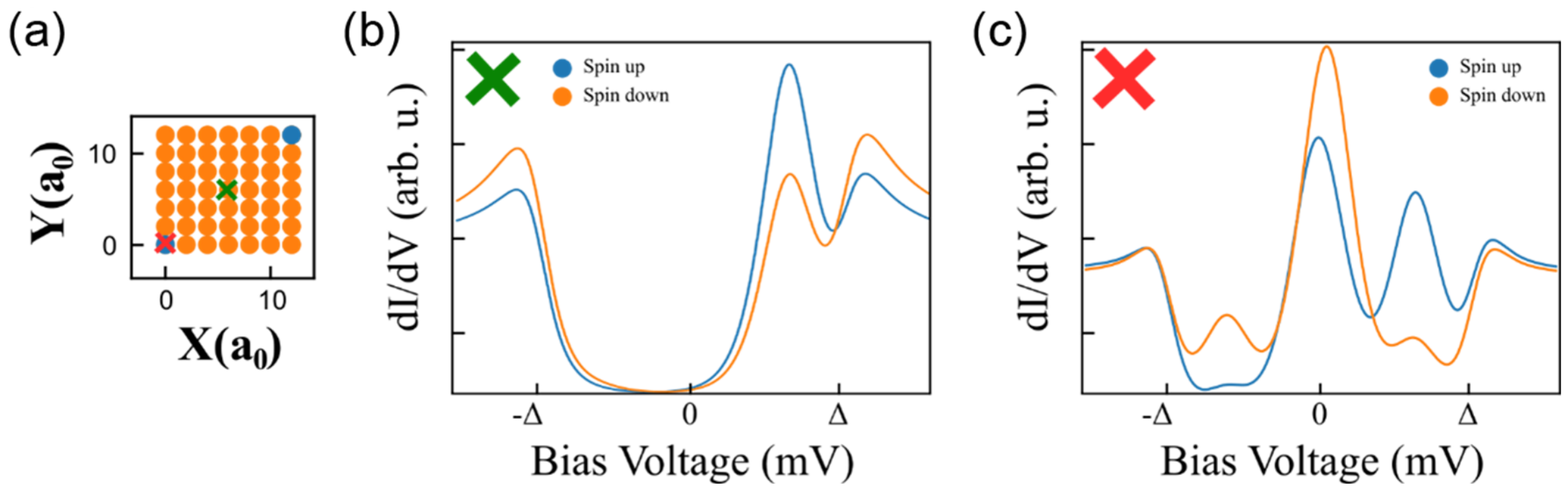

**Fig. S11| Spin polarization of a YSR dimer. a**, Distribution of two magnetic impurities. The green and red arrows are the positions where the dI/dV curves in panels **b** and **c** were calculated, respectively. **b** and **c**, Spin-polarized dI/dV curves (both polarizations) at the points marked with the green and red arrows in panel **a**, respectively.

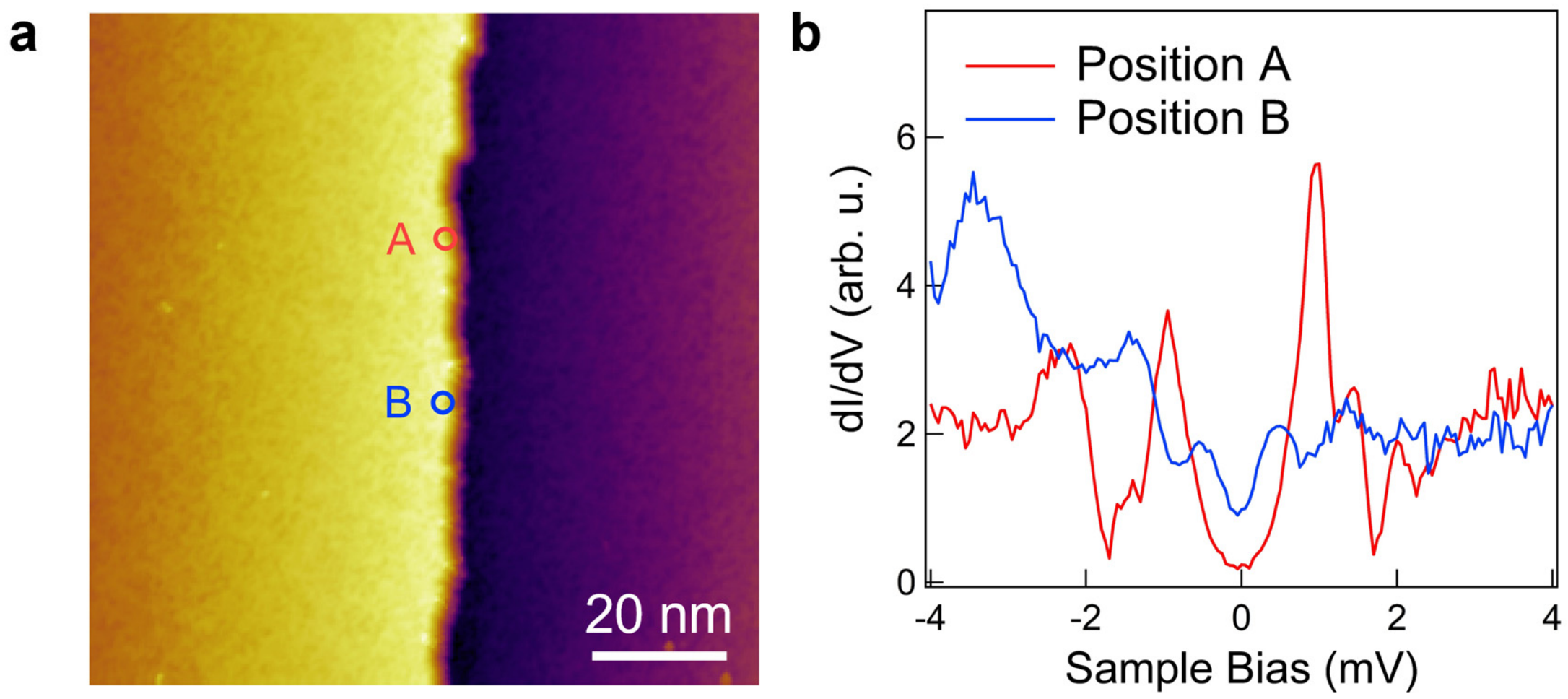

**Fig. S12| Two more STS on the step edge. a**, 100 nm × 100 nm STM topography image taken at $V =$ 0.5 V, $I$ =80 pA. **b**, Two dI/dV curves at the positions marked with A and B in **a** respectively. $V_{stab}$= 6 mV, $I_{stab}$= 80 pA, $V_{osc}$ = 0.03 mV.

## Supplementary Tables 1-2

| | $a$ | $b$ | $y_0$ | $A$ | $x_0$ | $B$ |
|---|---|---|---|---|---|---|
| Fig. S1b | -2224.6 | 0.01 | | | | |
| Fig. S1d | -1806.5 | 0.0011 | 0.0047 | 1573.1 | 681.56 | 1187.9 |
| Fig. S1e | -1802.7 | 0.0012 | -0.13 | 22074.2 | 581.53 | 15713.2 |
| Fig. S1g | -1834.8 | 0 | 0.053 | 246.94 | 233.46 | 130.78 |
| Fig. S1h | -1834.8 | 0.000343 | -0.44 | 858.18 | 261.79 | 463.86 |
| Fig. S5c | -1774.2 | 0.000071 | -0.041 | 284.23 | 178.26 | 120.25 |

**Table. S1|Fitting parameters used in Fig. S1 and Fig. S5.**

| | FWHM | Peak Max. (Å) | Bias (V) | Setpoint (pA) |
|---|---|---|---|---|
| Fig. S1d | 7.28 | 1.33 | -1.0 | 100 |
| Fig. S1e | 22.808 | 1.27 | -1.0 | 100 |
| Fig. S1g | 2.26 | 1.94 | -0.5 | 200 |
| Fig. S1h | 3.378 | 1.40 | -0.5 | 200 |
| Fig. S5c | 2.14 | 2.312 | 0.1 | 2000 |

**Table. S2|Extracted FWHM and peak height maximum in Fig. S1 and Fig. S5 together with the tunneling parameters.**

## Supplementary References

[1] A. Schlenhoff, S. Krause, G. Herzog, and R. Wiesendanger, Applied Physics Letters **97** (2010).
[2] A. Kubetzka, O. Pietzsch, M. Bode, and R. Wiesendanger, Physical Review B **67**, 020401 (2003).
[3] L. Yu, Acta Physica Sinica **21**, 75 (1965).
[4] H. Shiba, Prog. Theor. Phys. **40**, 435 (1968).
[5] A. I. Rusinov, Sov. J. Exp. Theor. Phys. **29**, 1101 (1969).
[6] J. Senkpiel, *et al.,* Physical Review B **100**, 014502 (2019).
[7] J. Ortuzar, *et al.*, Physical Review B **105**, 245403 (2022).
[8] S. Trivini, *et al.*, Physical Review B **110**, 235405 (2024).
[9] https://github.com/nanogunecic09/green-ysr.
[10] J. Senkpiel, *et al.*, Commun Phys **3**, 131 (2020).